\newcommand*{\ATLASLATEXPATH}{latex/}

\documentclass[cernpreprint,texlive=2011,txfonts,UKenglish]{latex/atlasdoc}
\pdfoutput=1

\usepackage[subfigure=true]{\ATLASLATEXPATH atlaspackage}
\usepackage{\ATLASLATEXPATH atlasbiblatex}

\usepackage{\ATLASLATEXPATH atlascontribute}

\usepackage{\ATLASLATEXPATH atlasphysics}
\graphicspath{{logos/}{figures/}}

\usepackage{VjjHinvPaper-defs}

\newcommand{\Hinv}{H\to \text{inv.}}

\newcommand\mjj{\ensuremath{m_{jj}}}

\def\met{\ensuremath{E_{\mathrm{T}}^{\mathrm{miss}}}}
\def\mpt{\ensuremath{p_{\mathrm{T}}^{\mathrm{miss}}}}

\def\metvec{\ensuremath{{\vec{E}_{\mathrm{T}}^{\mathrm{miss}}}}}
\def\mptvec{\ensuremath{{\vec{p}_{\mathrm{T}}^{\mathrm{miss}}}}}
\def\pt{\ensuremath{p_{\mathrm{T}}}}

\def\mt{\ensuremath{m_{\mathrm{T}}}}
\def\mtw{\ensuremath{m_{\mathrm{T}}^W}}

\def\vec#1{{\mbox{$\boldsymbol{#1}$}}}
\def\ifb{\ensuremath{\mathrm{fb}^{-1}}}

\def\ptV{\ensuremath{p^{V}_{\mathrm{T}}}}

\def\pt{\ensuremath{p_{\mathrm{T}}}}

\hypersetup{pdftitle={ATLAS draft},pdfauthor={The ATLAS Collaboration}}

\AtlasTitle{
Search for invisible decays of the Higgs boson
produced in association with a hadronically decaying vector boson
in $pp$ collisions at $\sqrt{s} = 8$ TeV with the ATLAS detector
}

\author{The ATLAS Collaboration}

\PreprintIdNumber{CERN-PH-EP-2015-061}

\AtlasJournal{EPJC}

\AtlasAbstract{
A search for Higgs boson decays to invisible particles is performed using 20.3~\ifb~of $pp$ collision data at a centre-of-mass energy of 8 TeV recorded by the ATLAS detector at the Large Hadron Collider.
The process considered is Higgs boson production in association with
a vector boson ($V$ = $W$ or $Z$) that decays hadronically,
resulting in events with two or more jets and large missing transverse momentum.
No excess of candidates is observed in the data over the background expectation. 
The results are used to constrain $VH$ production followed by $H$ decaying to invisible particles for the Higgs boson mass range $115<m_H<300$~GeV.
The 95\% confidence-level observed upper limit on
$\sigma_{VH} \times \text{BR}(\Hinv)$ varies
from 1.6~pb at 115~GeV to 0.13~pb at 300~GeV.
Assuming Standard Model production and including the $gg\to H$
contribution as signal, the results also lead to an observed upper
limit of 78\% at 95\% confidence level on the branching ratio of Higgs bosons decays to invisible particles at a mass of 125~GeV.
}

\AtlasCoverAnalysisTeam{Daniel B\"uscher(*), Philippe Calfayan(*), Jianming Qian(*), David Salek(*), Xiaohu Sun, Jike Wang(*)}

\AtlasCoverEdBoardMember{Giacomo Polesello~(chair)} 
\AtlasCoverEdBoardMember{Naoko Kanaya} 
\AtlasCoverEdBoardMember{Rachid Mazini} 
\AtlasCoverEdBoardMember{Nikolaos Rompotis}

\AtlasCoverEgroupEditors{atlas-higg-2014-07-editors@cern.ch}

\AtlasCoverEgroupEdBoard{atlas-higg-2014-07-editorial-board@cern.ch}

\begin{document}

\maketitle

\section{Introduction}
\label{sec:intro}
Since the discovery of a Higgs boson with a mass of approximately
125 GeV~\cite{atlas:2012obs,cms:2012obs} at the LHC in 2012,
the properties of this new particle have been studied extensively.
All results obtained so far~\cite{Aad:2013wqa,Aad:2013xqa,Aad:2014gsu,Aad:2015lxb,Khachatryan:2014mrj,Khachatryan:2014jba,Khachatryan:2014gse}
are consistent with the expectations
of the long-sought Standard Model (SM) Higgs
boson~\cite{Englert:1964aa,Higgs:1964aa,Higgs:1964ab,Guralnik:1964aa}.
However, sizeable deviations from the SM expectation
cannot be yet excluded; the total branching ratio of beyond-the-SM 
decays of the Higgs boson is only weakly constrained,
and its value could be as high as
$\sim 40\%$~\cite{Khachatryan:2014jba,Chatrchyan:2014tja}.
One possible decay is to weakly interacting particles, as predicted
by many extensions of the SM, e.g. Higgs boson portal models~\cite{Djouadi:1996mj,Martin:1999qf,Englert:2011yb,Djouadi:2011aa}.
In these models, the Higgs boson can decay to a pair of
dark-matter particles if kinematically allowed.
These decays are generally ``invisible'' to detectors, resulting
in events with large missing transverse momentum ($\met$).

Searches for Higgs boson decays to invisible particles ($\Hinv$)
have been performed by both the ATLAS and CMS collaborations~\cite{Aad:2014iia,Chatrchyan:2014tja}.
For example, the ATLAS Collaboration has placed an upper limit
of 75\%~\cite{Aad:2014iia} on the branching ratio of $\Hinv$ from 
Higgs boson production in association with a $Z$ boson
identified from its leptonic decays ($Z\to ee,\mu\mu$).
The present paper describes an independent search for the $\Hinv$ decay in final states
with two or more jets and large $\met$, motivated by Higgs boson production in
association with a vector boson $V$ ($V=W$ or $Z$): $q\bar{q}'\to VH$.
The vector boson is identified through its decay to a pair of quarks,
reconstructed as hadronic jets in the ATLAS detector, $V\to jj$.
Gluon fusion production $gg\to H$ followed by $\Hinv$ can also lead
to events with two or more jets and large $\met$, and therefore contributes 
to the signal of the search.
Negligible contributions of approximately 1\% and 0.2\% to the sensitivity come from 
$q\bar{q}'\to q\bar{q}'H $ production via vector-boson fusion (VBF) and from $qq/gg\to t\bar{t}H$ ($ttH$) production, respectively.
The VBF contribution is strongly suppressed by the $m_{jj}$
(dijet invariant mass) window cuts and by the forward-jet veto
used to reduce the top quark-antiquark background ($t\bar{t}$), as 
described in Sect.~\ref{sec:selection}.   
In a previous ATLAS dark-matter search,
limits on Higgs boson decays to invisible particles in $VH$ production were set
using events with a hadronically decaying vector boson and $\met$ as well~\cite{Aad:2013oja}.
However, the present analysis achieves better sensitivity
by using different techniques and performing dedicated optimizations.


\section{Experimental setup}
\label{sec:exp}

This search is based on proton--proton collision data at a centre-of-mass
energy of 8~TeV recorded 
with the ATLAS detector~\cite{Aad:2008zzm} in 2012, corresponding to
an integrated luminosity of 20.3~\ifb.
The ATLAS detector is a general-purpose detector with an inner
tracking system, electromagnetic and hadronic calorimeters, and 
a muon spectrometer surrounding the interaction
point.\footnote{The ATLAS experiment uses a right-handed coordinate
system with its origin at the nominal interaction point (IP) in the centre of the detector
and the $z$-axis along the beam pipe. The $x$-axis points from the IP to the centre of the LHC ring,
and the $y$-axis points upward. Cylindrical coordinates $(r,\phi)$ are used in the transverse
plane, $\phi$ being the azimuthal angle around the $z$-axis. The pseudorapidity is defined in
terms of the polar angle $\theta$ as $\eta=-\ln[\tan(\theta/2)]$. Transverse momenta are computed from
the three-momenta, $\vec{p}$, as $\pt = |\vec{p}| \sin\theta $.}
The inner tracking system is immersed in a 2~T axial magnetic field,
and the muon spectrometer employs a toroidal magnetic field.
Only data recorded when all subdetector systems were functional are used in this analysis.

The trigger system is organised in three levels. The first level is based on custom-made
hardware and uses coarse-granularity calorimeter and muon information. The second and
third levels are implemented as software algorithms and use the full detector granularity.
At the second level, only regions deemed interesting at the first level are analysed, while the
third level, called the event filter, makes use of the full detector read-out to reconstruct and
select events, which are then logged for offline analysis at a rate of up to 400 Hz averaged
over an accelerator fill.

\section{Object reconstruction and simulated samples}
\label{sec:objects}

Jets are reconstructed using the anti-$k_t$ algorithm~\cite{Cacciari:2008gp}
with a radius parameter of $R=0.4$.
Jet energies are corrected for the average contributions from 
minimum-bias interactions within the same bunch
crossing as the hard-scattering process and within
neighbouring bunch crossings (pile-up).
Furthermore, for jets with $\pt<50$~GeV and $|\eta|<2.4$,
the scalar sum of the $\pt$ of tracks matched to the jet and originating from
the primary vertex\footnote{The primary vertex is taken to be the reconstructed
vertex with the highest $\Sigma\pt^{2}$ of the associated tracks.}
must be at least 50\% of the scalar sum of the $\pt$ of all tracks matched to the jet,
to suppress jets from pile-up interactions.
Jets must have $\pt>20$~GeV ($\pt>30$~GeV) for $|\eta|<2.5$ ($2.5 <|\eta|< 4.5$).

Jets containing $b$-hadrons ($b$-jets) are identified ($b$-tagged) using the MV1c algorithm,
which is an improved version of the MV1 algorithm~\cite{ATLAS-CONF-2014-004}
with higher rejection of jets containing $c$-hadrons ($c$-jets).
It combines in a neural network the information from
various algorithms based on track impact-parameter significance or
explicit reconstruction of secondary decay vertices.
The operating point of this algorithm chosen for this analysis has an
efficiency of about 70\% for $b$-jets in $t\bar{t}$ events and
a $c$-jet (light-jet) mis-tag rate less than $20\%$ ($1\%$).

Lepton (electron or muon) candidates are identified in two categories:
loose and tight, in order of increasing purity.
Electron candidates are reconstructed from energy clusters in the electromagnetic calorimeter
matched to reconstructed tracks in the inner tracking system.
They are identified using likelihood-based methods~\cite{ATLAS-CONF-2014-032,Aad:2014fxa}.
Loose electrons must satisfy ``very loose likelihood'' identification criteria and are required
to have $\pt>7$~GeV and $|\eta|<2.47$.
Tight electrons are selected from the loose electrons and must also satisfy 
the ``very tight likelihood'' identification criteria.
Muon candidates are reconstructed using information from the inner tracker and the muon spectrometer~\cite{Aad:2014rra}.
Loose muons are required to have $\pt>7$~GeV and $|\eta|<2.7$.
Tight muons are then selected from the loose muons,
by requiring $\pt>25$~GeV and $|\eta|<2.5$.
They must be reconstructed in both the muon spectrometer and the inner tracker.
For the loose leptons, the scalar sum of the transverse momenta of
tracks within a cone of size $\Delta R  = \sqrt{(\Delta\phi)^2+(\Delta\eta)^2} = 0.2$ around the lepton candidate,
excluding its own track,
is required to be less than 10\% of the transverse momentum of the lepton.
For the tight leptons, there are more stringent isolation requirements:
the sum of the calorimeter energy deposits in a cone of size $\Delta R = 0.3$
around the lepton candidate, excluding the energy associated with it,
must be less than $4\%$ of the lepton candidate energy,
and the track-based isolation requirement is tightened from 10\% to 4\%.

The missing transverse momentum vector, $\metvec$, is computed
using fully calibrated and reconstructed physics objects,
as well as clusters of calorimeter-cell energy deposits that
are not associated with any object~\cite{ATLAS-CONF-2013-082}. 
Only calibrated jets with $\pt$ greater than 20 GeV are used in the
computation. The jet energy is also corrected for pile-up effects~\cite{Cacciari:2008gn}.
A track-based missing transverse momentum vector, $\mptvec$,
is calculated as the negative vector sum of transverse momenta
of reconstructed tracks associated with the primary vertex
and within $|\eta|<2.5$.  

Monte Carlo~(MC) simulated samples are produced for both the signal and
background processes.
Unless otherwise stated, the simulation~\cite{Aad:2010ah} is performed using the
ATLFAST-II package~\cite{ATL-PHYS-PUB-2010-013}, which combines a parameterized
simulation of the ATLAS calorimeter with the
{\sc Geant}4-based~\cite{Agostinelli:2002hh} full simulation for the
rest of the subdetector systems.

Signal events from $q\bar{q}'\to VH$ with $\Hinv$ are produced using
the NLO {\sc Powheg} method as implemented in the {\sc Herwig++} generator~\cite{Bahr:2008pv}. 
The $gg\to ZH$ production process contributes approximately 5\% to the total $ZH$ cross section.
Events from the $gg\to ZH$ production process are not simulated, but are taken into account by increasing
the $q\bar{q}\to ZH$ cross section as a function of the Higgs boson $\pt$ by the appropriate amount. 
The gluon-fusion signal events are produced using the {\sc Powheg} generator interfaced to {\sc Pythia8}
for parton showering and hadronization. The production of $q\bar{q}'\to VH$
followed by the SM $H\to b\bar{b}$ decay is considered as a background for the search.
The {\sc Pythia8} generator is used to produce these events.
The cross sections of all Higgs production processes are taken from Ref.~\cite{Heinemeyer:2013tqa}.

A significant source of background is the production of $V+$jets and of $t\bar{t}$ events.
A sample of $V+$jets events is generated using the {\sc Sherpa}
generator~\cite{Gleisberg:2008ta} with massive $b$- and $c$-quarks.
Events from the $t\bar{t}$ process are generated using the {\sc Powheg}
generator interfaced with {\sc Pythia}6~\cite{Sjostrand:2006za}. 
Other background contributions include diboson ($WW, WZ\ {\rm and}\ ZZ$) and
single top-quark production.
The {\sc Powheg} generator interfaced to {\sc Pythia}8 is used to produce diboson events.
The diboson cross sections are calculated at NLO in QCD using the MCFM
program~\cite{Campbell:2010ff} with the MSTW2008NLO parton distribution functions (PDFs)~\cite{Martin:2009iq}.
The $s$-channel and $Wt$ single top-quark events are produced using the {\sc Powheg} generator,
as for $t\bar{t}$ production. The remaining $t$-channel process is simulated with the
{\sc AcerMC} generator~\cite{Kersevan:2004yg} interfaced to {\sc Pythia}6.
Cross sections of the three single top-quark processes are taken from
Refs.~\cite{Kidonakis:2011wy,Kidonakis:2010tc,Kidonakis:2010ux}.
Table~\ref{tab:MCs} summarizes the MC generators, PDFs and normalization cross sections used in this analysis.

\begin{table}
\caption{List of MC generators, parton distribution functions (PDFs)
and cross sections used for the signal and background processes.
The $\Hinv$ signal cross sections are given for $m_H = 125$~GeV and
assume SM production and BR($\Hinv$) = 100\%.
Details are given in the text.}
\centering
\begin{tabular}{lllc}
\hline
\hline
Process			& \multicolumn{1}{c}{Generator}		& PDFs				& Cross section [pb]		\\ \hline
\textbf{$t\bar{t}$}	& {\sc Powheg} + {\sc Pythia}		& CT10~\cite{Lai:2010vv}	& normalized to data		\\
\textbf{$V$+jets}       & {\sc Sherpa}				& CT10				& normalized to data		\\ \hline

\textbf{Single top}	&					&				&				\\
$t$-channel		& AcerMC				& CTEQ6L1~\cite{Pumplin:2002vw} &	88\phantom{.00000}	\\ 
$s$-channel		& {\sc Powheg} + {\sc Pythia}		& CT10				&	5.6\phantom{000}	\\ 
$Wt$			& {\sc Powheg} + {\sc Pythia}		& CT10				&	22\phantom{.00000}	\\ 
\hline 
\textbf{Diboson}	&					&				&				\\
$WW$			& {\sc Powheg} + {\sc Pythia}		& CT10				&	52\phantom{.00000}	\\ 
$WZ$			& {\sc Powheg} + {\sc Pythia} 		& CT10				&	9.2\phantom{000}	\\ 
$ZZ$			& {\sc Powheg} + {\sc Pythia}		& CT10				&	3.3\phantom{000}	\\ 
\hline
\textbf{SM VH}     	& 					&				&				\\
$q\bar{q}'\rightarrow VH(\rightarrow b\bar{b})$ & {\sc Pythia}  & CTEQ6L1			&	0.18\phantom{00}	\\
$gg\phantom{'}\rightarrow ZH(\rightarrow b\bar{b})$	& {\sc Powheg} + {\sc Pythia}	& CT10	&	0.0038			\\
\hline
\textbf{Signals}	&					&				&					\\
$q\bar{q}\phantom{'}\to Z(\rightarrow jj) H(\rightarrow \text{inv.})$  & {\sc Herwig++}		& CT10	&	0.29\phantom{00}	\\
$q\bar{q}'\to W(\rightarrow jj) H(\rightarrow \text{inv.})$  & {\sc Herwig++}			& CT10	&	0.48\phantom{00}	\\
$gg\phantom{'}\to  H(\rightarrow \text{inv.})$               & {\sc Powheg} + {\sc Pythia}	& CT10	&	19\phantom{.00000}	\\ 
\hline
\hline
\end{tabular}
\label{tab:MCs}
\end{table}

\section{Event selection}
\label{sec:selection}

Events are required to pass an $\met$ trigger with a threshold of 80~GeV, which is a cut applied at the third level.
The $\met$ trigger is fully efficient for $\met>160$~GeV and 97\% efficient for
$\met=120$~GeV. An efficiency correction is derived from $W\rightarrow \mu\nu$+jets and $Z\rightarrow\mu^{+}\mu^{-}$+jets events. 
This correction is below 1\% for 120~GeV$<\met<160$~GeV.
Events are also required to have $\met>120$~GeV,
$\mpt>30$~GeV, no loose leptons and
two or three ``signal jets''
(satisfying $|\eta|<2.5$, $\pt>20$~GeV and leading jet $\pt>45$~GeV).
The inclusion of 3-jet events improves the signal efficiency.
A requirement is made on $\HT$, defined as the scalar sum of the $\pt$ of all jets: $\HT>120\ (150)$~GeV for events with two (three) jets.
This cut is employed to avoid a trigger bias introduced by the dependence
of the trigger efficiency on the jet activity, as also discussed in Ref.~\cite{Aad:2014xzb}.
Events are discarded if they have additional jets with $\pt>20~(30)$~GeV and
$|\eta|<2.5$ ($2.5<|\eta|<4.5$) to reduce the contribution from the
$t\bar{t}$ background process.

For $VH$ signal events, $\met$ resulting from the $\Hinv$ decay is expected to be strongly correlated with the transverse momentum of the vector boson $V$ ($\pt^V$). 
Since the $\met$ distribution of the signal is harder than that of the background, 
additional sensitivity in the analysis is gained by optimizing the selection cuts separately for four $\met$ ranges. 
Here and in the following, the dijet refers to the two leading jets in events with three jets. 
The dijet invariant mass, $m_{jj}$, is required to be consistent with that of the $W/Z$ boson. 
In addition a requirement on the radial separation between the two jets, $\Delta R_{jj}$, 
is made as the jets are expected to be close in for highly boosted $V$-bosons. 
Both the $m_{jj}$ and the $\Delta R_{jj}$ cuts reduce the $V$+jets and the $t\bar{t}$ backgrounds, and depend on $\met$. 
The cut values are given in Table~\ref{tab:selections}.

Multijet events are copiously produced in hadron collisions.
Fluctuations in
jet energy measurements in the calorimeters can create $\met$ in these events
and therefore mimic the signal.
To suppress their contribution, additional selection criteria are applied to
the azimuthal angles between $\metvec$, $\mptvec$ and jets:
$\Delta\phi(\metvec,\mptvec)<\pi/2$, $\min[\Delta\phi(\metvec,{\rm jet})]>1.5$
and $\Delta\phi(\metvec,{\rm dijet})>2.8$.
Here $\Delta\phi(\metvec,\mptvec)$ is the azimuthal angle between $\metvec$ and
$\mptvec$, $\min[\Delta\phi(\metvec,{\rm jet})]$ the angle between $\metvec$ and
its nearest jet, and $\Delta\phi(\metvec,{\rm dijet})$ is the angle between
$\metvec$ and the momentum vector of the dijet system.
These requirements are based on characteristics of events with mismeasured $\met$ in the multijet background,
while taking advantage of the expected topologies of signal events. 

Finally, the selected events are further categorized according to $b$-tag
multiplicity (zero, one and two $b$-tagged jets) to improve the sensitivity.
Combined with the two categories in jet multiplicity (two and three jets),
there are in total six categories in the signal region.

\begin{table}[h]
\caption{The $\met$-dependent event selections of the signal region for the four $\met$ ranges.                                                 } 
\centering
\begin{tabular}{lcccc} 
\hline
\hline	
\MET\ range [\GeV]                    & 120--160 & 160--200 &  200--300 & $>$ 300         \\
\hline
Variable                        & \multicolumn{4}{c}{Selection} \\ \hline
$\Delta R_{jj}$, 2- and 3-jet events  & 0.7--2.0 & 0.7--1.5 & $<$ 1.0  & $<$ 0.9        \\
$m_{jj}$, 2-jet events [\GeV]	& 70--100 & 70--100 & 70--100 & 75--100  \\ 
$m_{jj}$, 3-jet events [\GeV]	& 50--100 & 55--100 & 60--100 & 70--100 \\ 
\hline
\hline
\end{tabular}
\label{tab:selections}
\end{table}

\section{Background estimation}
\label{sec:BG}
In addition to the signal region, a number of control regions,
designed to estimate various background contributions, are defined.
They include the signal sideband (events not passing the $m_{jj}$ requirement),
and the regions dominated by $V$+jets and $t\bar{t}$ events 
as discussed below.  
The multijet background is estimated from the data.
The distributions of the $V$+jets and $t\bar{t}$ backgrounds are taken from MC simulation
while their normalizations are estimated from the data.
The remaining diboson, single-top and SM VH(bb) backgrounds are obtained from MC simulation.

The multijet background is estimated using four regions defined
by requirements on $\Delta\phi(\metvec,\mptvec)$
and $\min[\Delta\phi(\metvec,{\rm jet})]$,
as listed in Table~\ref{tab:0LQCD}.
The shapes of the $m_{jj}$ and $\met$ distributions
in the signal region A are taken from region C and
the normalizations are determined by the ratio of
the numbers of events in regions B and D.

\begin{table}[h]
\caption{Definition of the signal region, A, and the three regions B, C and D used to estimate the multijet background in the signal region.}
\centering
\begin{tabular}{lcccc}
\hline
\hline
Region          & A & B & C & D         \\
\hline
$\Delta\phi(\metvec,\mptvec)$ & $< \pi/2$ & $< \pi/2$ & $> \pi/2$ & $> \pi/2$ \\
$\min[\Delta\phi(\metvec,{\rm jet})]$ & $> 1.5$ & $< 0.4$ & $> 1.5$ & $< 0.4$ \\
\hline
\hline
\end{tabular}
\label{tab:0LQCD}
\end{table}

The normalizations of the $V$+jets backgrounds are estimated using control regions enhanced in $W$+jets and $Z$+jets events. 
In all cases at least one lepton is required to have $\pt>$25~GeV.
The $W$+jets events are selected by requiring exactly one tight lepton,
$\met>20$~GeV ($\met>50$~GeV if $\pt^W>200$~GeV), exactly two signal jets
and $\mt^W<120$~GeV.\footnote{
The transverse mass, $\mt^W$, is calculated from the transverse momentum and the azimuthal
angle of the charged lepton, $\pt^\ell$ and $\phi^\ell$, and from the missing transverse momentum's 
magnitude, \met, and azimuthal angle, $\phi^{\mathrm{miss}}$:
$\mtw = \sqrt{2\pt^\ell\met (1-\cos(\phi^\ell-\phi^{\mathrm{miss}}))}$.
The transverse momentum of the $W$ boson, $\pt^W$, is reconstructed as the
magnitude of the vector sum of the lepton transverse momentum
and the \metvec.}
Moreover, $\pt^W>100$~GeV is required in order to approximately match
the phase space of the signal region.
The $Z$+jets events are selected by requiring two loose leptons of the same flavour with opposite charges with 
invariant mass $83<m_{\ell\ell}<99$~GeV, at least two signal jets and a dilepton
transverse momentum greater than 100~GeV. The kinematic distributions of the $V$+jets backgrounds
are obtained from simulation that takes into account the different flavour composition of the jets.
The simulated events are reweighted depending on the $\Delta\phi(\text{jet}_1,\text{jet}_2)$ and
$\pt^V$ to better match the data distributions~\cite{Aad:2014xzb}. 
The $Z$+jets control region has a small contribution from $t\bar{t}$ (1.3\%), which is
estimated using a $t\bar{t}$ control region.
This region is selected by
requiring events to have two oppositely charged leptons of different flavour (one of which has $\pt>$25 GeV) and passing the loose selection requirements, 
and at least two signal jets which are $b$-tagged.
The signal sideband and the $V$+jets control regions are divided to match the categorization of the signal region 
while the $t\bar{t}$ control region remains as one category as described above.
For the $V$+jets and $t\bar{t}$ control regions,
the distributions of the multijet background are obtained from control regions
defined by inverting the lepton isolation requirement
and the normalizations are determined by template fits~\cite{Aad:2014xzb}.

\section{Systematic uncertainties}
\label{sec:sys}
The experimental systematic uncertainties considered include the
trigger efficiency, object reconstruction and identification efficiency,
and object energy and momentum scales as well as resolutions.
Among these, the jet energy scale~(JES) and resolution~(JER) uncertainties
have the largest impact on the result.
The JES uncertainties are $\pm$3\% and $\pm$1\% for central jets with
a $\pt$ of 20 GeV and 1 TeV, respectively.
The JER uncertainty varies from between $\pm$10\% and $\pm$20\%,
depending on the pseudorapidities of the jets, for jets with $\pt=$ 20 GeV
to less than $\pm$5\% for jets with $\pt >$ 200~GeV.
The JER and JES uncertainties are also propagated to the $\met$ uncertainty.
The $b$-tagging uncertainty depends on jet $\pt$ and comes mainly from
the uncertainty on the measurement of the efficiency in $t\bar{t}$
events~\cite{ATLAS-CONF-2014-004}.
The dominant contribution arises from
jets matched to $b$-hadrons in the MC record of the particles' true identities.
Their efficiency uncertainties are at the level of $\pm$2--3\% over
most of the jet \pt\ range, but reach $\pm$5\% for $\pt=$ 20 GeV
and $\pm$8\% above $\pt=$ 200 GeV~\cite{Aad:2014bia}.
The uncertainty on the integrated luminosity is $\pm$2.8\%. It is derived following the same methodology as that detailed in Ref.~\cite{Aad:2013ucp}.

For the backgrounds, a large number of modelling systematic
uncertainties are considered, which account for possible differences between
the data and the MC models.
These uncertainties are estimated following the studies of
Ref.~\cite{Aad:2014xzb} and are briefly summarized here.
The uncertainties on the $V$+jets backgrounds come mainly from the
knowledge of jet flavour composition and the $\pt^V$, $\Delta\phi_{jj}$
and $\mjj$ distributions. 
For $t\bar{t}$ production, uncertainties on the top quark transverse
momentum and the $m_{jj}$, \met\ and $\pt^V$ distributions are considered.
The diboson background uncertainties are dominated by the theoretical
uncertainties of the cross-section predictions, which include contributions
from the renormalization and factorization scales and the choice of PDFs.
The robustness of the multijet background estimation is assessed by
varying the definition of the control regions B and D and
an uncertainty of $\pm$100\% is assigned for this small background
($<1$\% in the signal regions). 

The uncertainty on the signal acceptance is evaluated by changing the
factorization and renormalization scale parameters, parton distribution
function choices and the parton shower choices.
For the $VH$ signal, the dominant uncertainty is from parton shower
modelling, which can be as large as $\pm$8\%.
For the $gg\to H$ signal, the dominant uncertainty originates from
the renormalization and factorization scales
and can be as large as $\pm$15\% in the high \met\ regions.
Additional corrections to the Higgs boson $\pt$ distribution of the
$gg\to H$ signal are applied to match the distribution from a calculation at
NNLO+NNLL provided by HRes2.1~\cite{deFlorian:2012mx,Grazzini:2013mca}. The detailed precedures are following the ones 
used in the $H\rightarrow\gamma\gamma$ and $H\rightarrow WW^{*}$ analyses as described in Ref.~\cite{ATLAS:2014aga,Aad:2014eha}.
The related uncertainties are also taken into account.

\section{Results}
\label{sec:results}

\begin{table}
\caption{Predicted and observed numbers of events for the six categories in the signal region.
The yields and uncertainties of the backgrounds are shown after the profile likelihood fit
to the data.
In this fit all categories share the same signal-strength parameter.
The quoted uncertainties combine the statistical and systematic contributions. 
These can be smaller for the total background than for
individual components due to anti-correlations. 
The yields and uncertainties
of the signals are shown as expected before the fit
for $m_H=125$~GeV and BR$(\Hinv)=100\%$. 
Signal contributions from VBF and $t\bar{t}H$ production are estimated to be negligible.}
\centering
\begin{tabular}{llrrrrrrrr}\hline\hline												
 & $b$-tag category	& \multicolumn{2}{c}{0-tag}	&\hspace*{0.8cm}& \multicolumn{2}{c}{1-tag}	&\hspace*{0.8cm}& \multicolumn{2}{c}{2-tag}		\\ \hline
 & Process & \multicolumn{8}{c}{2-jet events} \\ \hline 													
Background & $Z$+jets	&	24400	& $\pm$ \phantom{.}1100	&&	1960	& $\pm$	\phantom{.}200	&&	164	& $\pm$	\phantom{.0}13		\\
 	& $W$+jets	&	20900	& $\pm$ \phantom{.0}770	&&	1160	& $\pm$	\phantom{.}130	&&	47	& $\pm$	\phantom{.00}7	\\
 	& $\ttbar$	&	403	& $\pm$ \phantom{.00}74	&&	343	& $\pm$	\phantom{.0}65	&&	57	& $\pm$	\phantom{.0}10		\\
 	& Single top	&	149	& $\pm$ \phantom{.00}16	&&	107	& $\pm$	\phantom{.0}14	&&	11	& $\pm$	\phantom{.00}2	\\
 	& Diboson	&	1670	& $\pm$ \phantom{.0}180	&&	227	& $\pm$	\phantom{.0}25	&&	64	& $\pm$	\phantom{.00}7	\\
 	& SM VH($bb$)	&	1.5	& $\pm$ \phantom{00}0.5	&&	6	& $\pm$	\phantom{.00}2	&&	3	& $\pm$	\phantom{.00}1	\\
 	& Multijet	&	26	& $\pm$ \phantom{.00}43	&&	8	& $\pm$	\phantom{.00}7	&&	0.7	& $\pm$	\phantom{0}0.9	\\
Total  &		&	47560	& $\pm$ \phantom{.0}490	&&	3804	& $\pm$	\phantom{.0}64	&&	347	& $\pm$	\phantom{.0}15	\\
\hline													
Signal & $gg\rightarrow H$    &	403	& $\pm$ \phantom{.00}95	&&	25	& $\pm$	\phantom{.00}6	&&	2.1	& $\pm$	\phantom{0}0.5	\\
 & $W(\rightarrow jj) H$	&	425	& $\pm$ \phantom{.00}45	&&	44	& $\pm$	\phantom{.00}6	&&	0.6	& $\pm$	\phantom{0}0.1	\\
 & $Z(\rightarrow jj) H$	&	217	& $\pm$ \phantom{.00}19	&&	42	& $\pm$	\phantom{.00}4	&&	26	& $\pm$	\phantom{.00}2	\\ \hline
Data  &	& \multicolumn{2}{c}{47404}	&& \multicolumn{2}{c}{3831}	&& \multicolumn{2}{c}{344}		\\ \hline \\
													
 & & \multicolumn{8}{c}{3-jet events} \\ \hline													
Background & $Z$+jets	&	9610	& $\pm$ \phantom{.0}580	&&	795	& $\pm$	\phantom{.0}93	&&	53	& $\pm$	\phantom{.00}7	\\
 	& $W$+jets	&	7940	& $\pm$ \phantom{.0}510	&&	479	& $\pm$	\phantom{.0}70	&&	21	& $\pm$	\phantom{.00}4	\\
 	& $\ttbar$	&	443	& $\pm$ \phantom{.00}53	&&	437	& $\pm$	\phantom{.0}53	&&	63	& $\pm$	\phantom{.00}7	\\
 	& Single top	&	97	& $\pm$ \phantom{.00}14	&&	66	& $\pm$	\phantom{.00}9	&&	6.4	& $\pm$	\phantom{0}0.9	\\
 	& Diboson	&	473	& $\pm$ \phantom{.00}54	&&	55	& $\pm$	\phantom{.00}6	&&	13	& $\pm$	\phantom{.00}2	\\
 	& SM VH($bb$)	&	0.8	& $\pm$ \phantom{00}0.3	&&	2.6	& $\pm$	\phantom{0}0.9	&&	1.4	& $\pm$	\phantom{0}0.5	\\
 	& Multijet	&	22	& $\pm$ \phantom{.00}29	&&	4	& $\pm$	\phantom{.00}4	&&	0.6	& $\pm$	\phantom{0}0.6	\\
Total  	&		&	18580	& $\pm$ \phantom{.0}200	&&	1840	& $\pm$	\phantom{.0}40	&&	158	& $\pm$	\phantom{.00}7	\\
\hline													
Signal & $gg\rightarrow H$                &	224	& $\pm$	\phantom{.00}55	&&	15	& $\pm$	\phantom{.00}4	&&	1.2	& $\pm$	\phantom{0}0.5	\\
 & $W(\rightarrow jj) H$	&	110	& $\pm$ \phantom{.00}16	&&	11	& $\pm$	\phantom{.00}1	&&	0.14	& $\pm$	\phantom{}0.03	\\
 & $Z(\rightarrow jj) H$       	&	65	& $\pm$ \phantom{.000}7	&&	12	& $\pm$	\phantom{.00}1	&&	6.1	& $\pm$	\phantom{0}0.7	\\ \hline
Data &	& \multicolumn{2}{c}{18442}	&& \multicolumn{2}{c}{1842}	&& \multicolumn{2}{c}{159}			\\
\hline\hline													
\end{tabular}
\label{tab:event_yields}
\end{table}

The potential $\Hinv$ signal is extracted through a combined likelihood
fit to the observed $\met$ distributions of the signal region and its
sideband and the $\pt^V$ distributions of the control regions
($\pt^V$ is defined as $\pt^W$, $\pt^Z$ and $\pt^{e+\mu}$ for the $W$+jets,
$Z$+jets and $t\bar{t}$ control regions, respectively). The normalizations of the $V$+jets and $t\bar{t}$ 
backgrounds are free parameters in this fit.   
The $\met$ distributions are binned in such a way that each bin yields
approximately the same amount of expected signal.
The 2-jet categories of the signal region are split into ten bins,
while fewer bins are used in the 3-jet categories and the sideband.
Most $V$+jets control regions are split into five $\ptV$ bins,
each yielding approximately the same amount of expected background.
The 0-tag category of the $V$+jets control regions and the $t\bar{t}$
control region are used inclusively in the fit.
The signal strength $\mu$, defined as the ratio of the signal yield
($\sigma_{VH} \times \text{BR}(\Hinv)$)
relative to the SM production cross section and assuming
BR$(\Hinv)=100\%$, is used to parameterize the signal in the data.
A binned likelihood function is constructed as the product of Poisson
probability terms comparing the numbers of events observed in the data
to those expected from the assumed signals and estimated background
contributions for all categories of the signal and control regions.
The likelihood takes into account the background normalization and the
systematic uncertainties.
It is maximized to extract the most probable signal-strength value, $\hat{\mu}$.

Table~\ref{tab:event_yields} shows the numbers of observed events in
the data compared to the numbers of estimated background events from
the likelihood fit for each signal category.
In all categories the data agrees with the background estimation. The backgrounds are dominated by $Z$+jets and $W$+jets events. 
Subleading backgrounds come from top and diboson production. The SM $VH$ and multijet background contributions are very small with the final event selection.

The fit reveals no significant excess of events over the background
expectations and yields a best-fit signal-strength value of
$\hat{\mu} = -0.13^{+0.43}_{-0.44}$, which is consistent with zero.
The contributions from the individual systematic uncertainties are summarized in Table~\ref{tab:muData}.
The systematic uncertainty sources which have the largest impacts are
the energy scale of the jets and of $\met$ along with the
modelling (shape and normalization) of the diboson and $V$+jets backgrounds.
The \MET\ distributions of the events passing the signal region selection
are shown in Figs.~\ref{fig:postfit_SRs_2j} and~\ref{fig:postfit_SRs_3j}
after the profile likelihood fit to the data. 
The fit results are also propagated to the $m_{jj}$ distributions
of the events passing the signal region selection (without the $m_{jj}$-window cuts).
The corresponding plots are shown in Figs.~\ref{fig:postfit_mass_0tag},
\ref{fig:postfit_mass_1tag}~and~\ref{fig:postfit_mass_2tag}
for the three $b$-tag categories separately.

\begin{table}[htbp]
\caption{
Impacts of sources of systematic uncertainty on
the uncertainty of the fitted signal strength, $\Delta\mu$,
in the data. Only sources with contributions larger than $\pm 0.03$ are listed.}
\label{tab:muData}
\begin{center}
\begin{tabular}{lll} \hline\hline
 Source        & \multicolumn{2}{c}{Impact on $\Delta\mu$}  \\ \hline
  \multicolumn{3}{c}{Object systematic uncertainties} \\ \hline
  Jets \& \MET  & $+$0.22 & $-$0.22\\
  Luminosity   & $+$0.04  & $-$0.03 \\
  $b$-tagging  & $+$0.05 & $-$0.04 \\  \hline
  \multicolumn{3}{c}{Background systematic uncertainties} \\ \hline
  Diboson      & $+$0.26 & $-$0.29\\
  $Z$+jets     & $+$0.21 & $-$0.22\\
  $W$+jets     & $+$0.15 & $-$0.16\\
  $t\bar{t}$   & $+$0.06 & $-$0.05\\
  Multijet     & $+$0.07 & $-$0.07 \\   \hline
  \multicolumn{3}{c}{Total} \\ \hline
  Total systematic uncertainty & $+$0.41 & $-$0.43\\
  Data statistical uncertainty & $+$0.12 & $-$0.12\\ \hline
  Total uncertainty & $+$0.43 & $-$0.44\\ \hline\hline
\end{tabular}
\end{center}
\end{table}

\begin{figure*}
\centering
\subfigure[]{
	\includegraphics[width=0.48\textwidth]{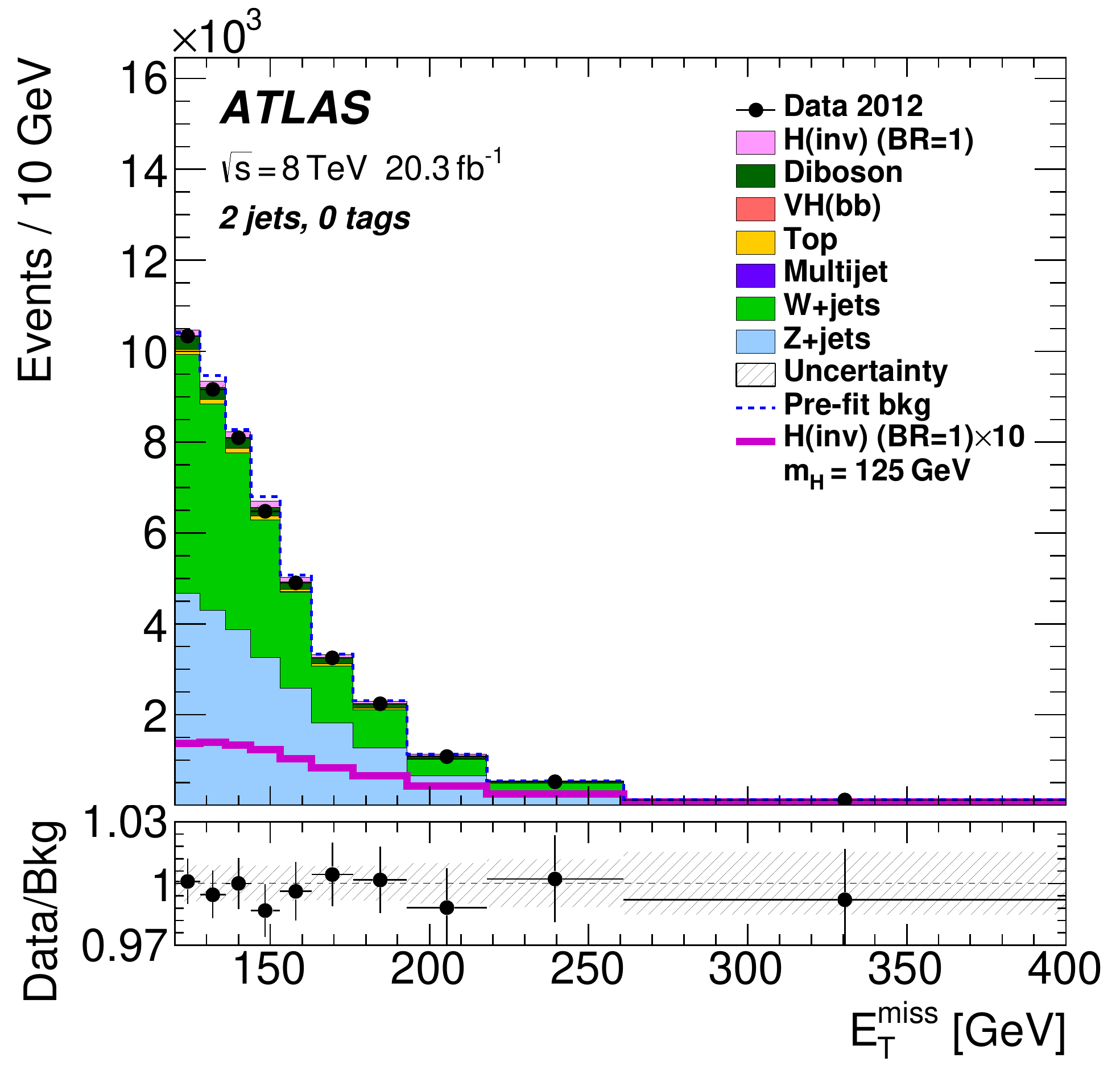}
}
\subfigure[]{
	\includegraphics[width=0.48\textwidth]{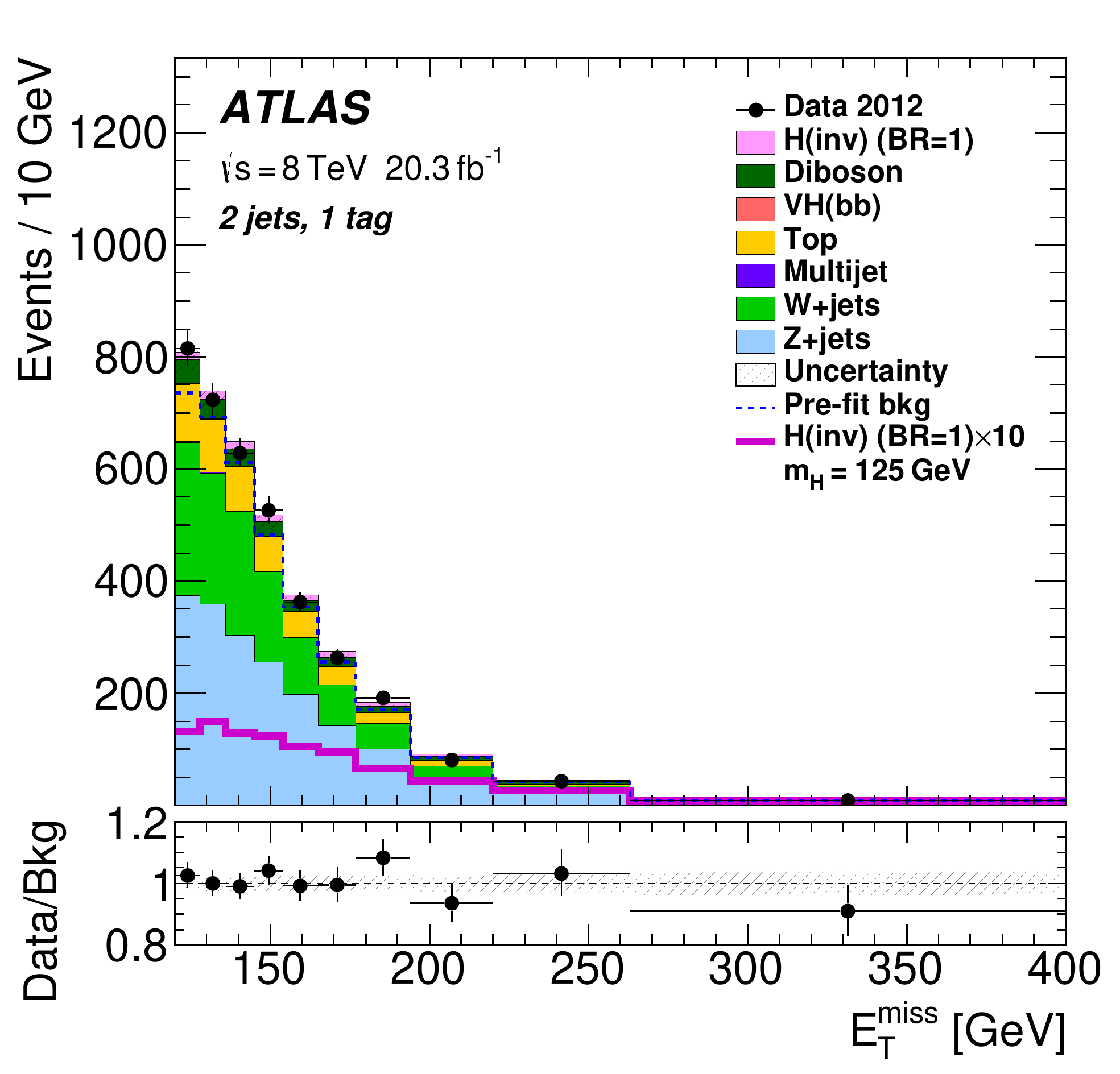}
}
\subfigure[]{
	\includegraphics[width=0.48\textwidth]{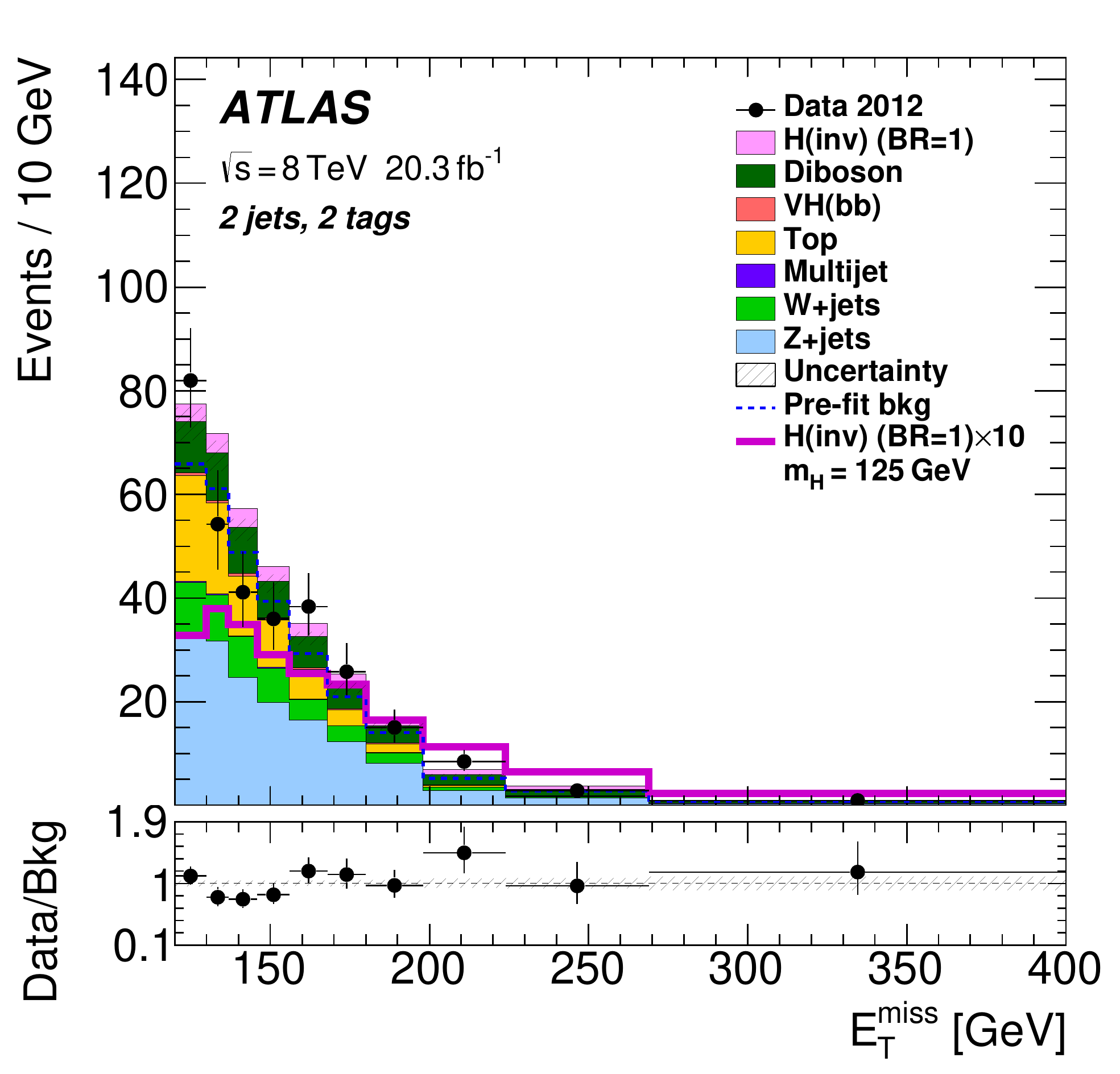}
}
\caption{
The missing transverse momentum ($\met$) distributions of the 2-jet events in
the signal region for the (a) 0-$b$-tag, (b) 1-$b$-tag and (c) 2-$b$-tag categories.
The data are compared with the background model after the likelihood fit.
The bottom plots show the ratio of the data to the total background.
The signal expectation for $m_H = 125\,$GeV and BR$(\Hinv)=100\%$ is shown on top of the
background and additionally as an overlay line, scaled by the factor
indicated in the legend.
The total background before the fit is shown as a dashed line.
The hatched bands represent the total uncertainty on the background.
}
\label{fig:postfit_SRs_2j}
\end{figure*}

\begin{figure*}
\centering
\subfigure[]{
	\includegraphics[width=0.48\textwidth]{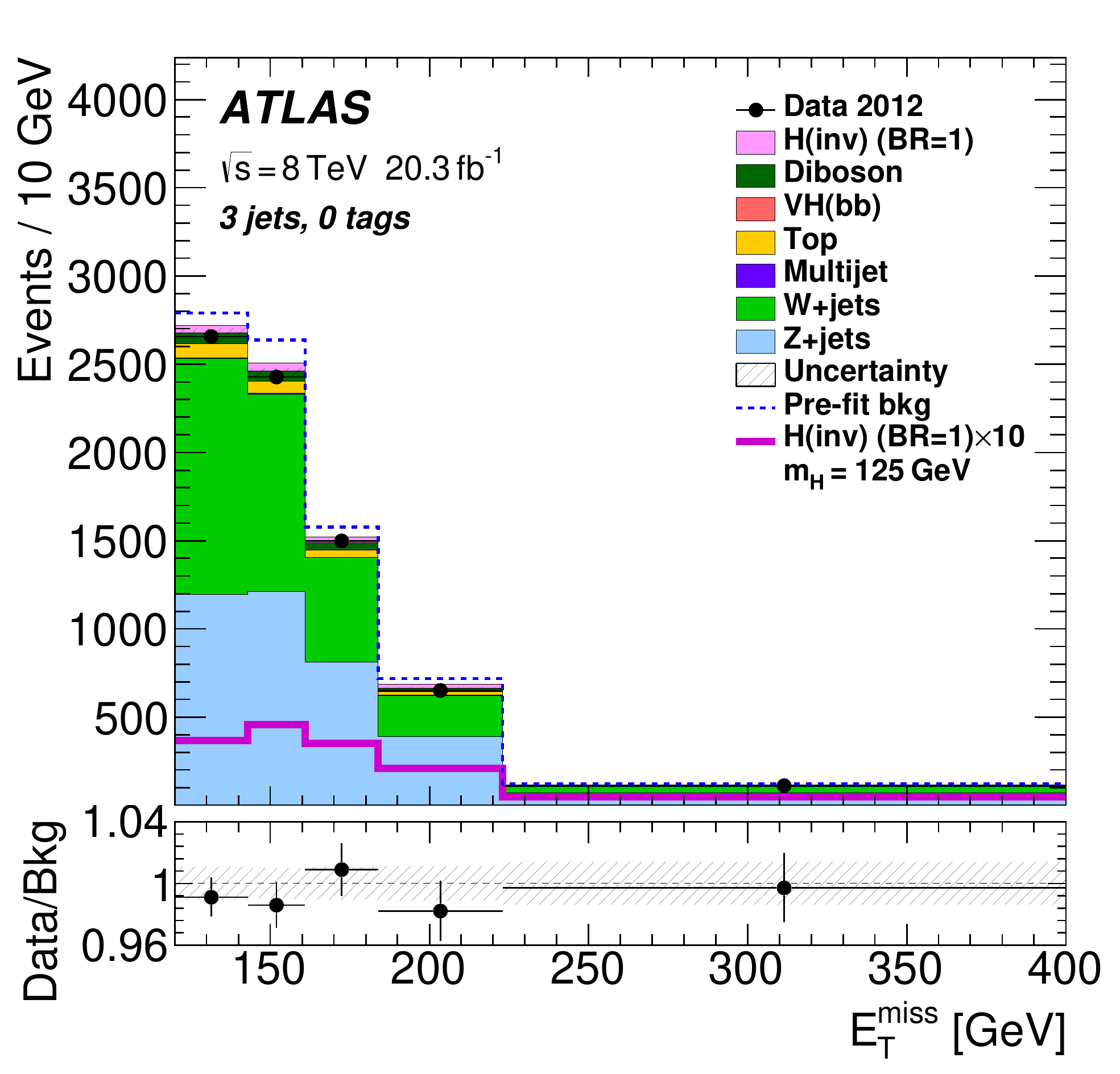}
}
\subfigure[]{
	\includegraphics[width=0.48\textwidth]{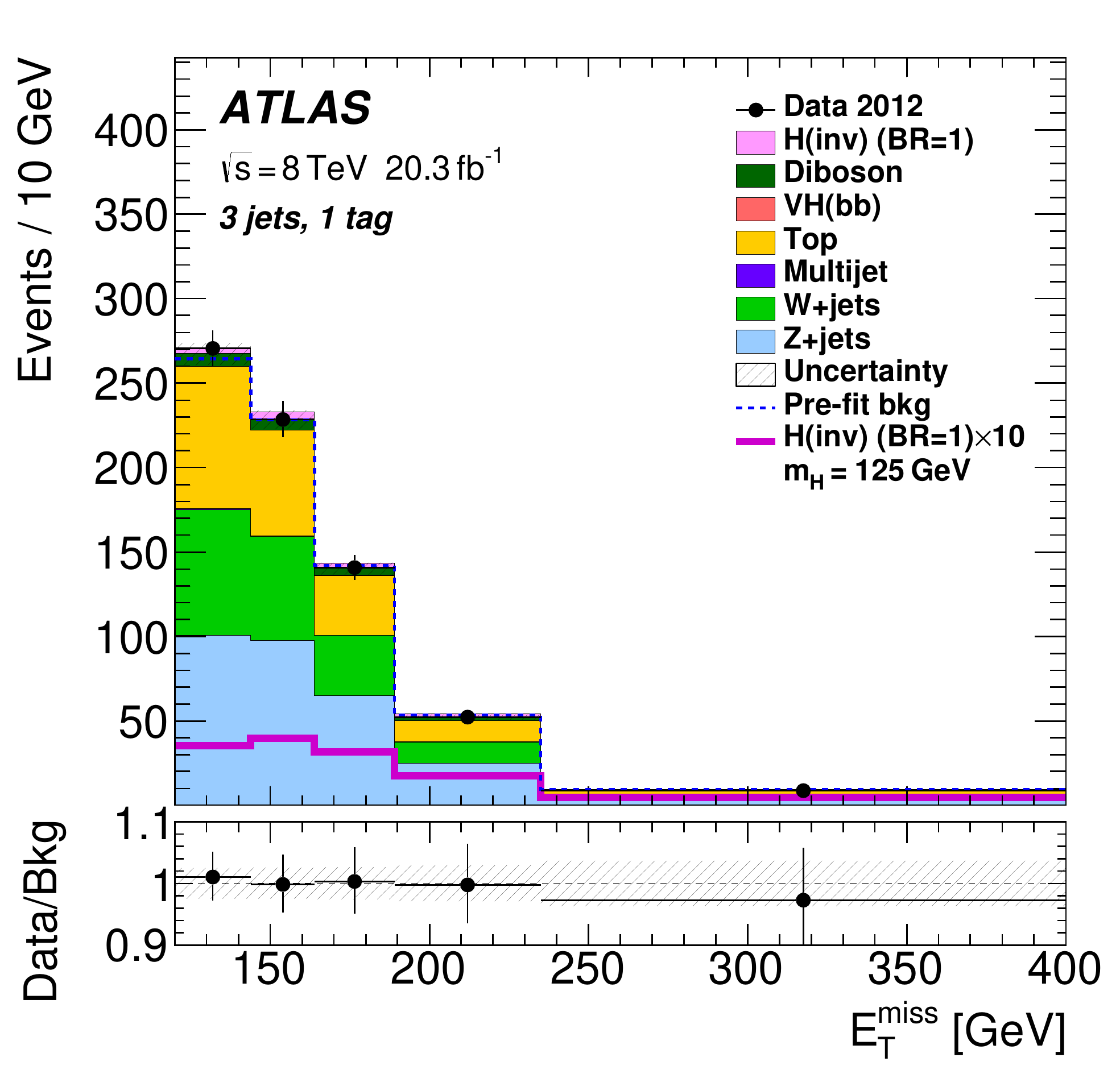}
}
\subfigure[]{
	\includegraphics[width=0.48\textwidth]{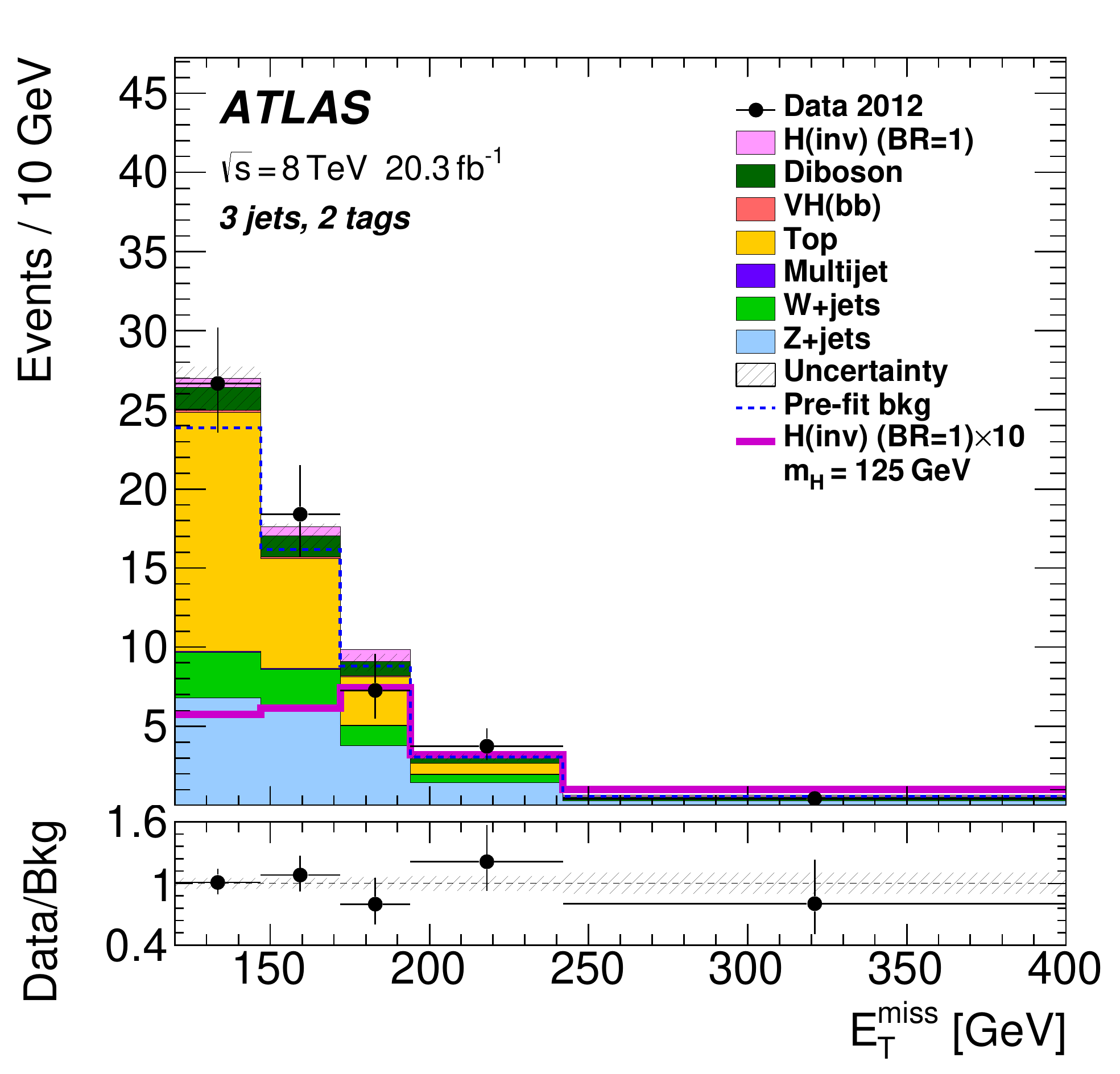}
}
\caption{
The missing transverse momentum ($\met$) distributions of the 3-jet events in
the signal region for the (a) 0-$b$-tag, (b) 1-$b$-tag and (c) 2-$b$-tag categories.
The data are compared with the background model after the likelihood fit.
The bottom plots show the ratio of the data to the total background.
The signal expectation for $m_H = 125\,$GeV is shown on top of the
background and additionally as an overlay line, scaled by the factor
indicated in the legend.
The total background before the fit is shown as a dashed line.
The hatched bands represent the total uncertainty on the background.
}
\label{fig:postfit_SRs_3j}
\end{figure*}

\begin{figure*}
\centering
\subfigure[]{
	\includegraphics[width=0.48\textwidth]{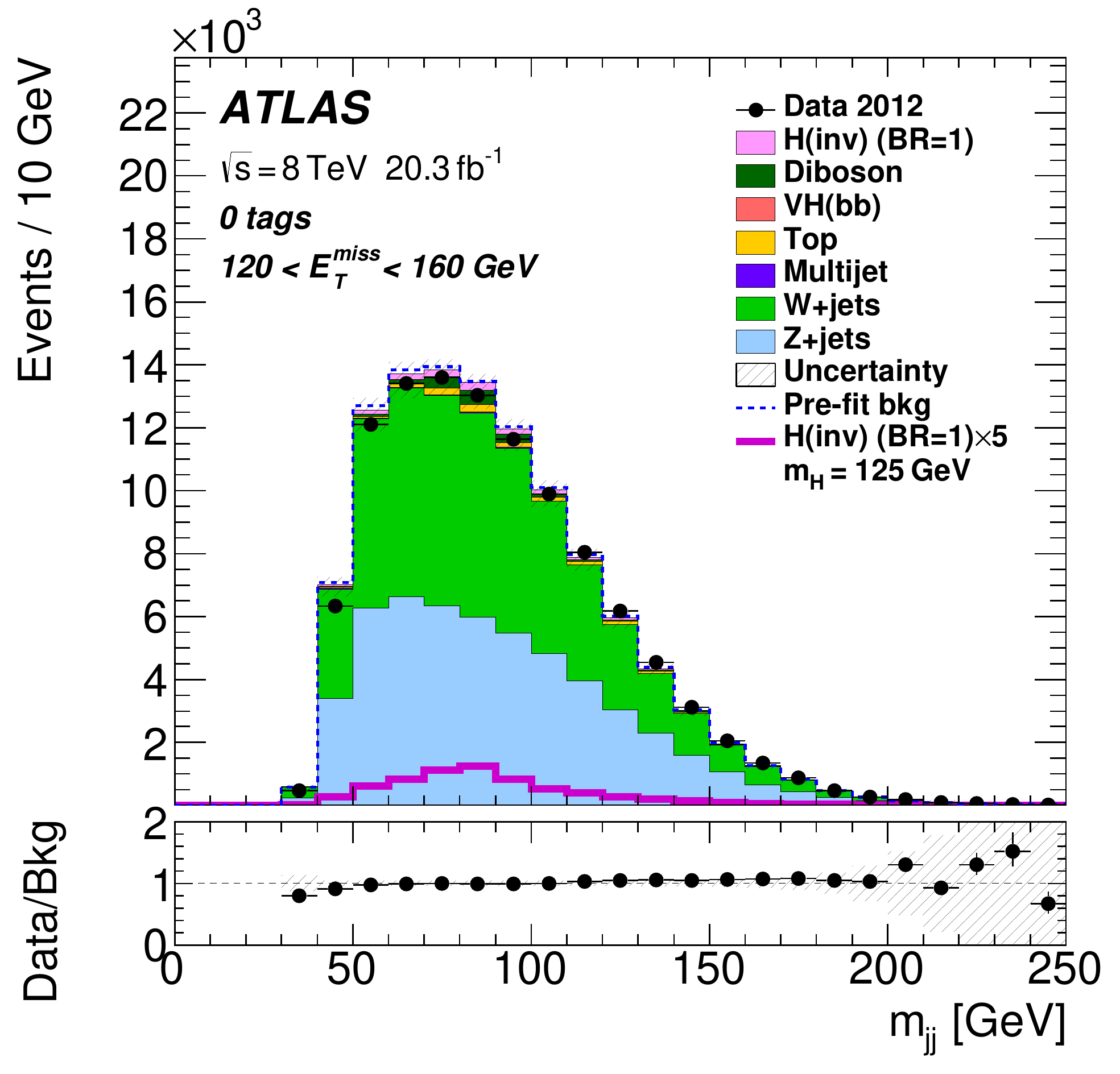}
}
\subfigure[]{
	\includegraphics[width=0.48\textwidth]{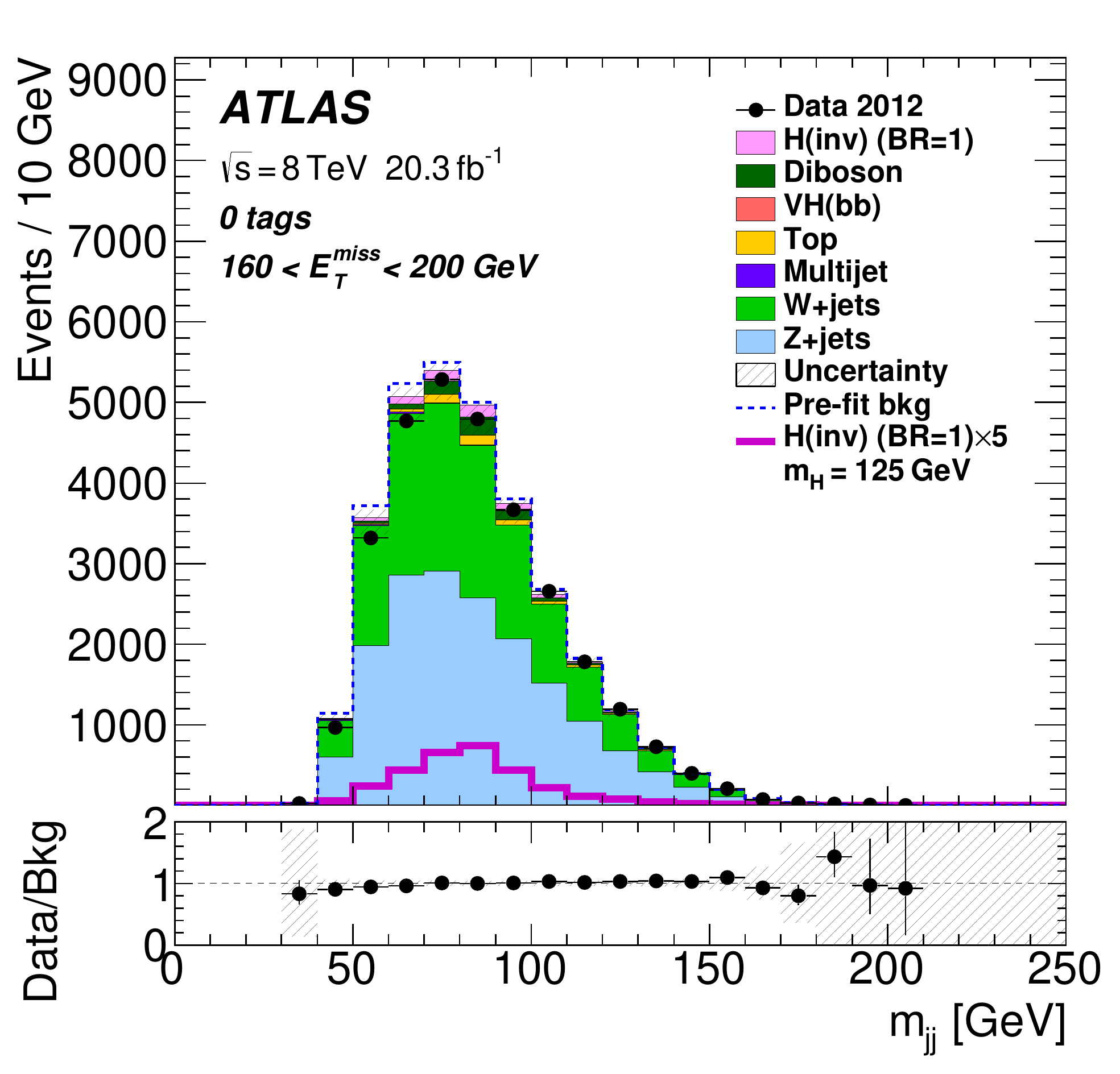}
}
\subfigure[]{
	\includegraphics[width=0.48\textwidth]{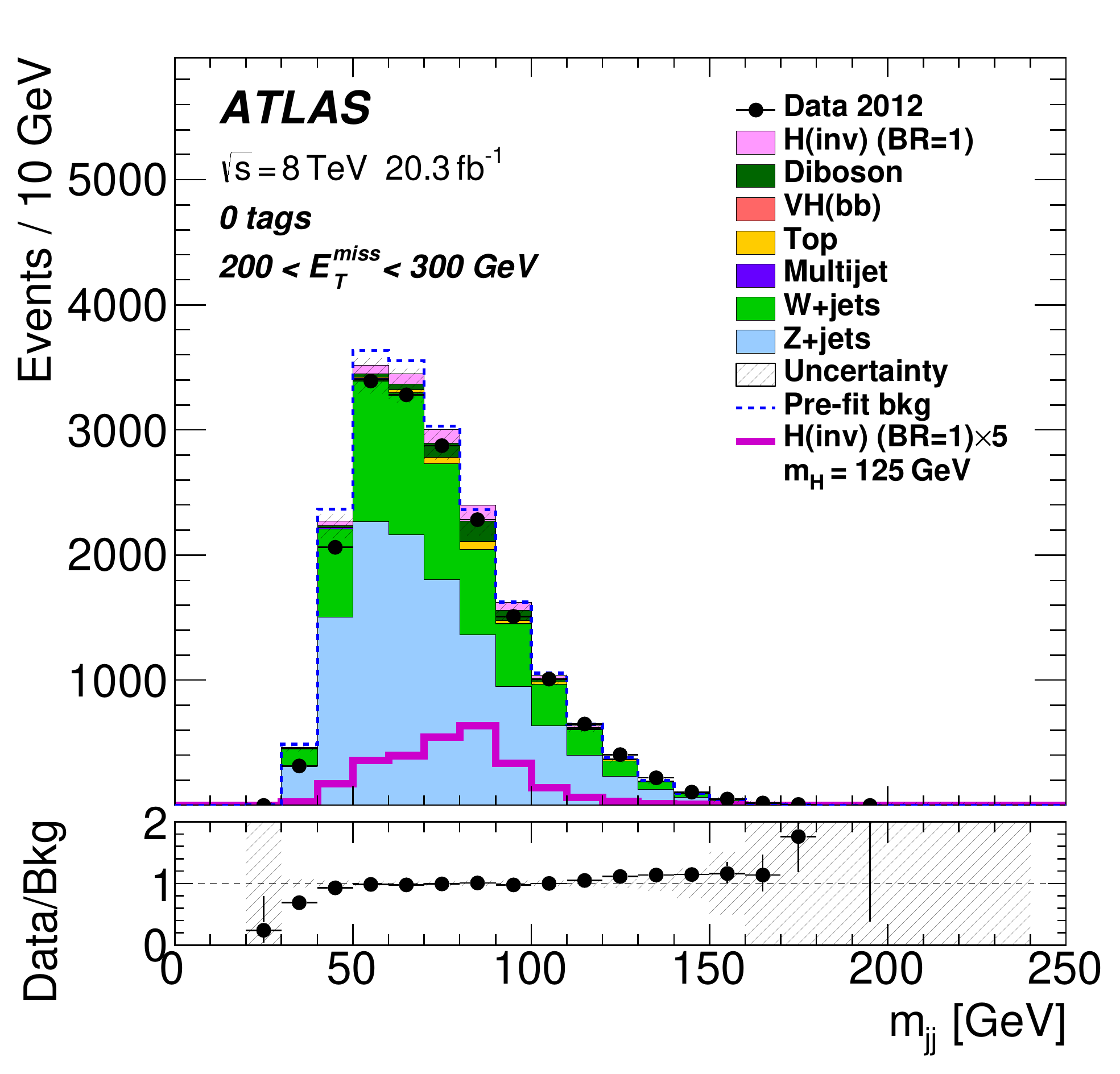}
}
\subfigure[]{
	\includegraphics[width=0.48\textwidth]{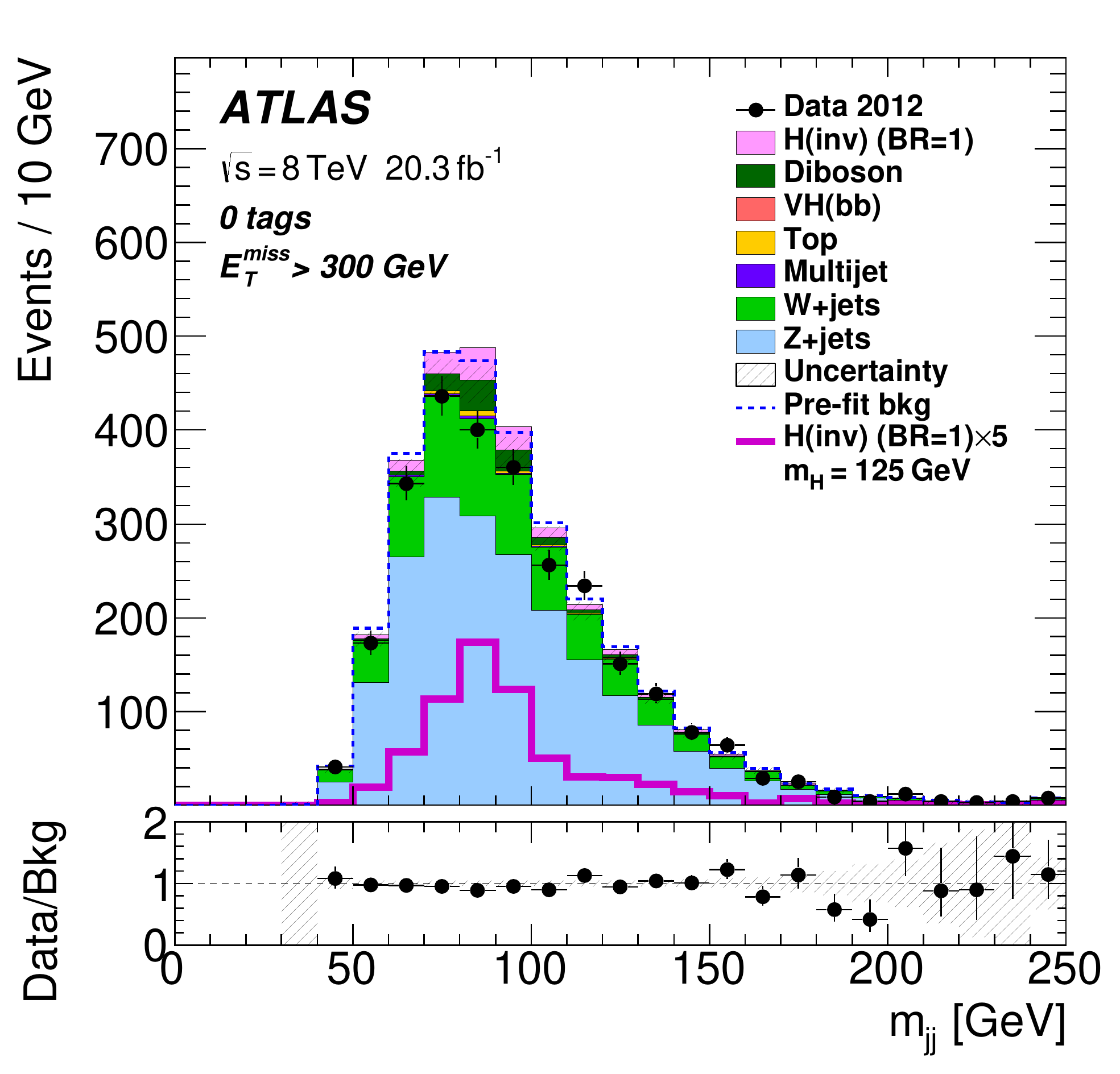}
}
\caption{
The dijet invariant mass ($m_{jj}$) distributions in
the signal region for the 0-$b$-tag category, for events with \met\ in the range
(a) [120--160 GeV], (b) [160--200 GeV], (c) [200--300 GeV] and (d) [$>$ 300 GeV].
The data are compared with the background model after the likelihood fit.
The bottom plots show the ratio of the data to the total background.
The signal expectation for $m_H = 125\,$GeV is shown on top of the
background and additionally as an overlay line, scaled by the factor
indicated in the legend.
The total background before the fit is shown as a dashed line.
The hatched bands represent the total uncertainty on the background.
}
\label{fig:postfit_mass_0tag}
\end{figure*}

\begin{figure*}
\centering
\subfigure[]{
	\includegraphics[width=0.48\textwidth]{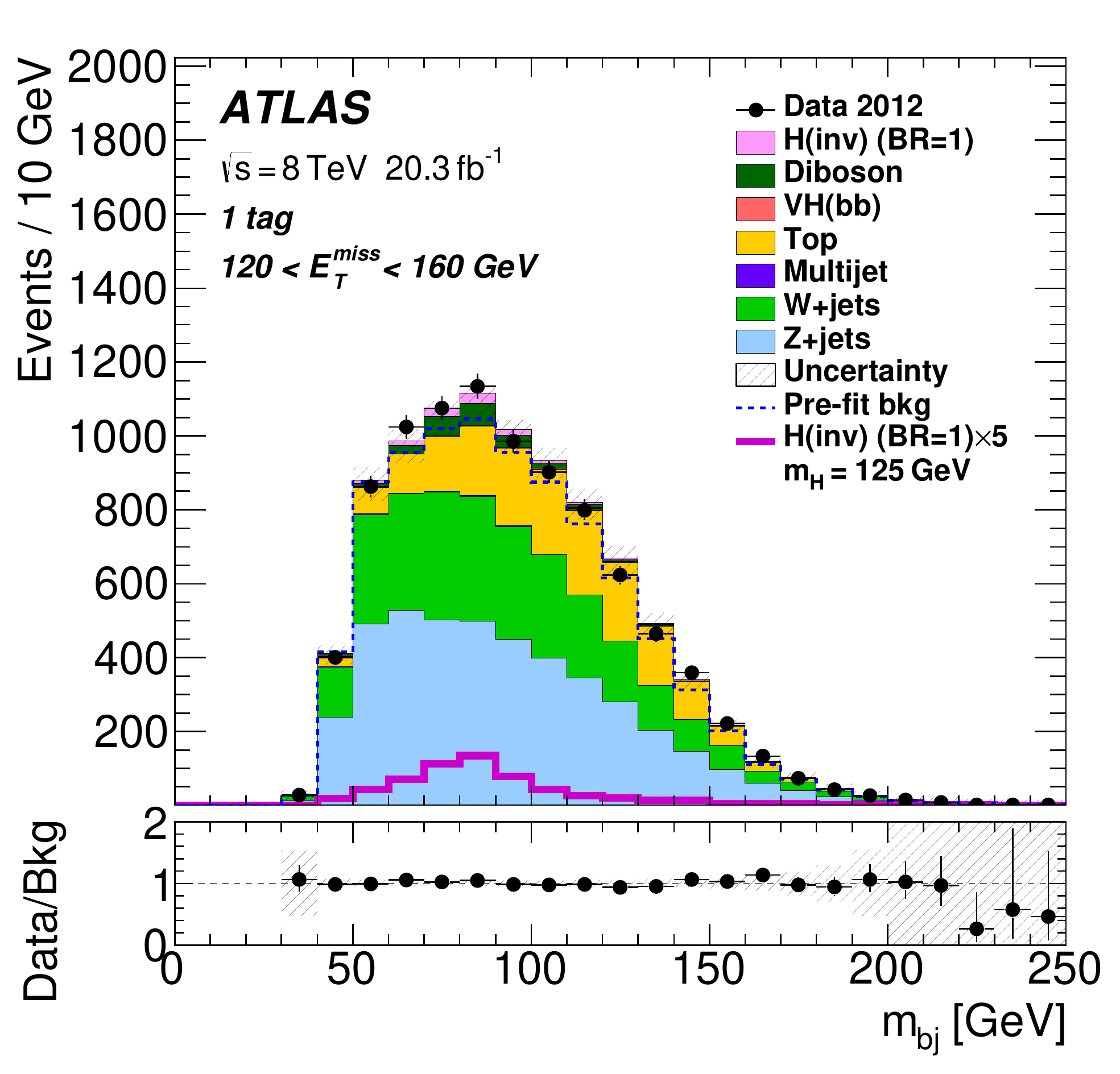}
}
\subfigure[]{
	\includegraphics[width=0.48\textwidth]{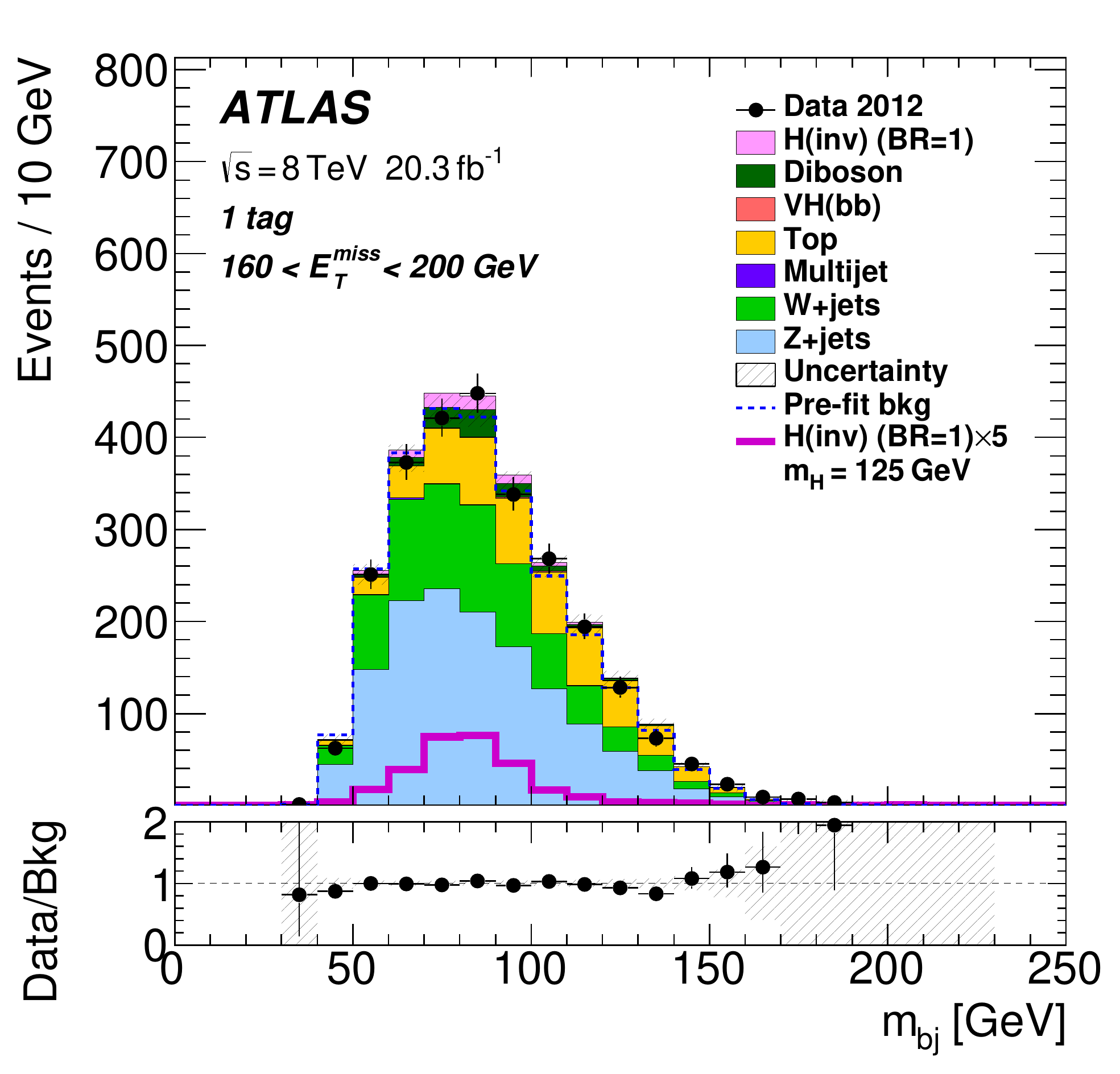}
}
\subfigure[]{
	\includegraphics[width=0.48\textwidth]{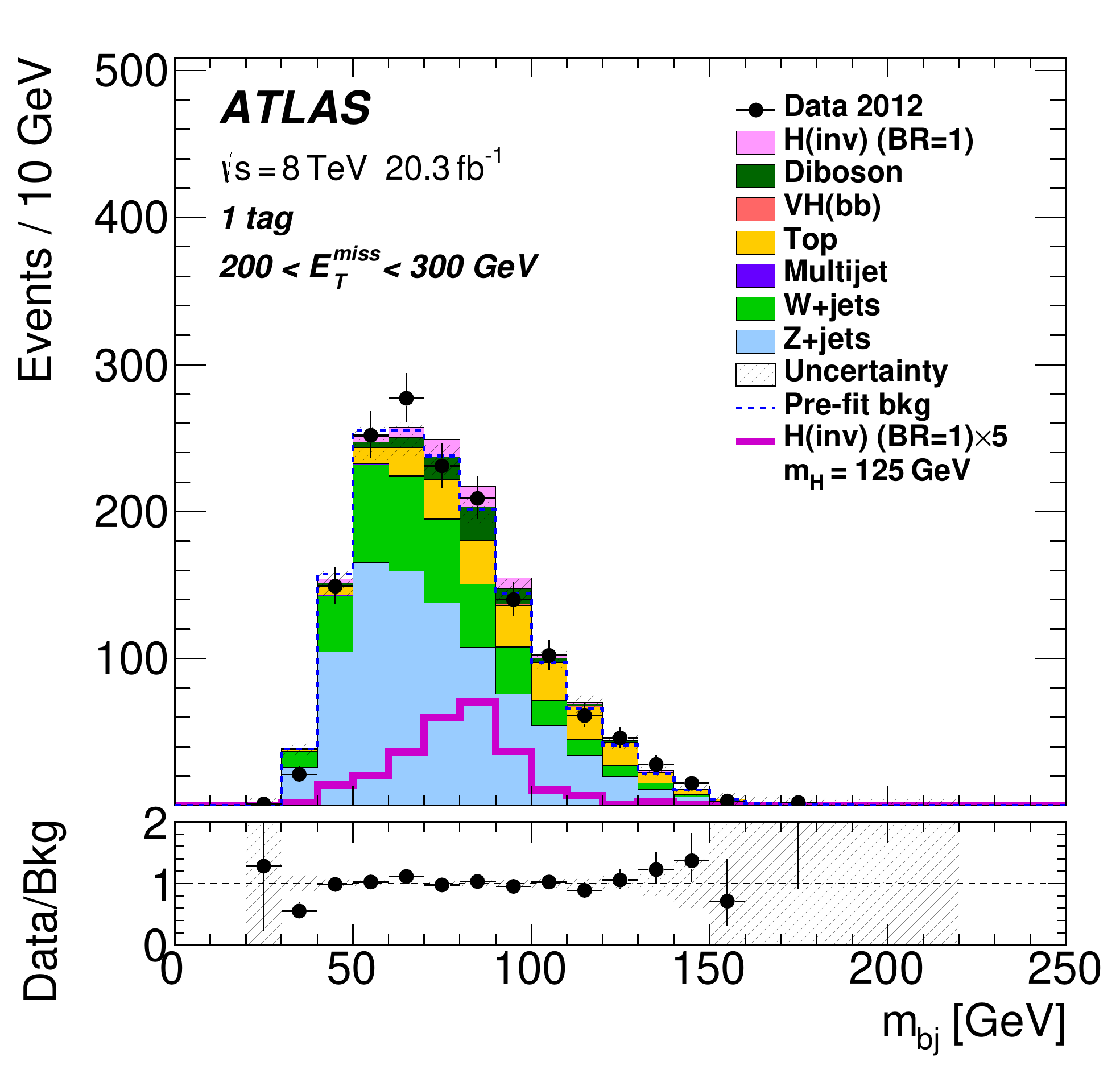}
}
\subfigure[]{
	\includegraphics[width=0.48\textwidth]{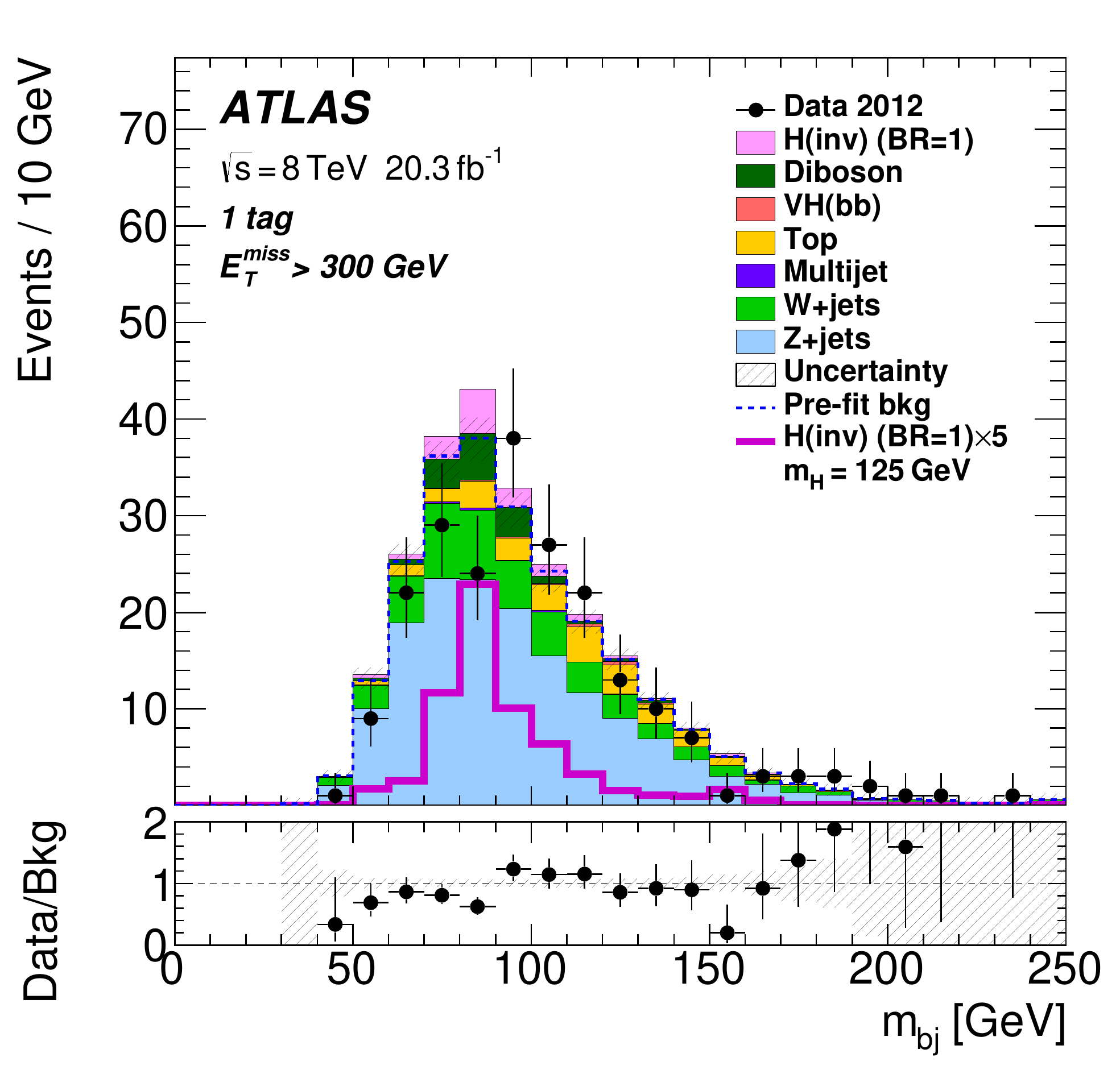}
}
\caption{
The dijet invariant mass ($m_{bj}$) distributions in
the signal region for the 1-$b$-tag category, for events with \met\ in the range
(a) [120--160 GeV], (b) [160--200 GeV], (c) [200--300 GeV] and (d) [$>$ 300 GeV].
The data are compared with the background model after the likelihood fit.
The bottom plots show the ratio of the data to the total background.
The signal expectation for $m_H = 125\,$GeV is shown on top of the
background and additionally as an overlay line, scaled by the factor
indicated in the legend.
The total background before the fit is shown as a dashed line.
The hatched bands represent the total uncertainty on the background.
}
\label{fig:postfit_mass_1tag}
\end{figure*}

\begin{figure*}
\centering
\subfigure[]{
	\includegraphics[width=0.48\textwidth]{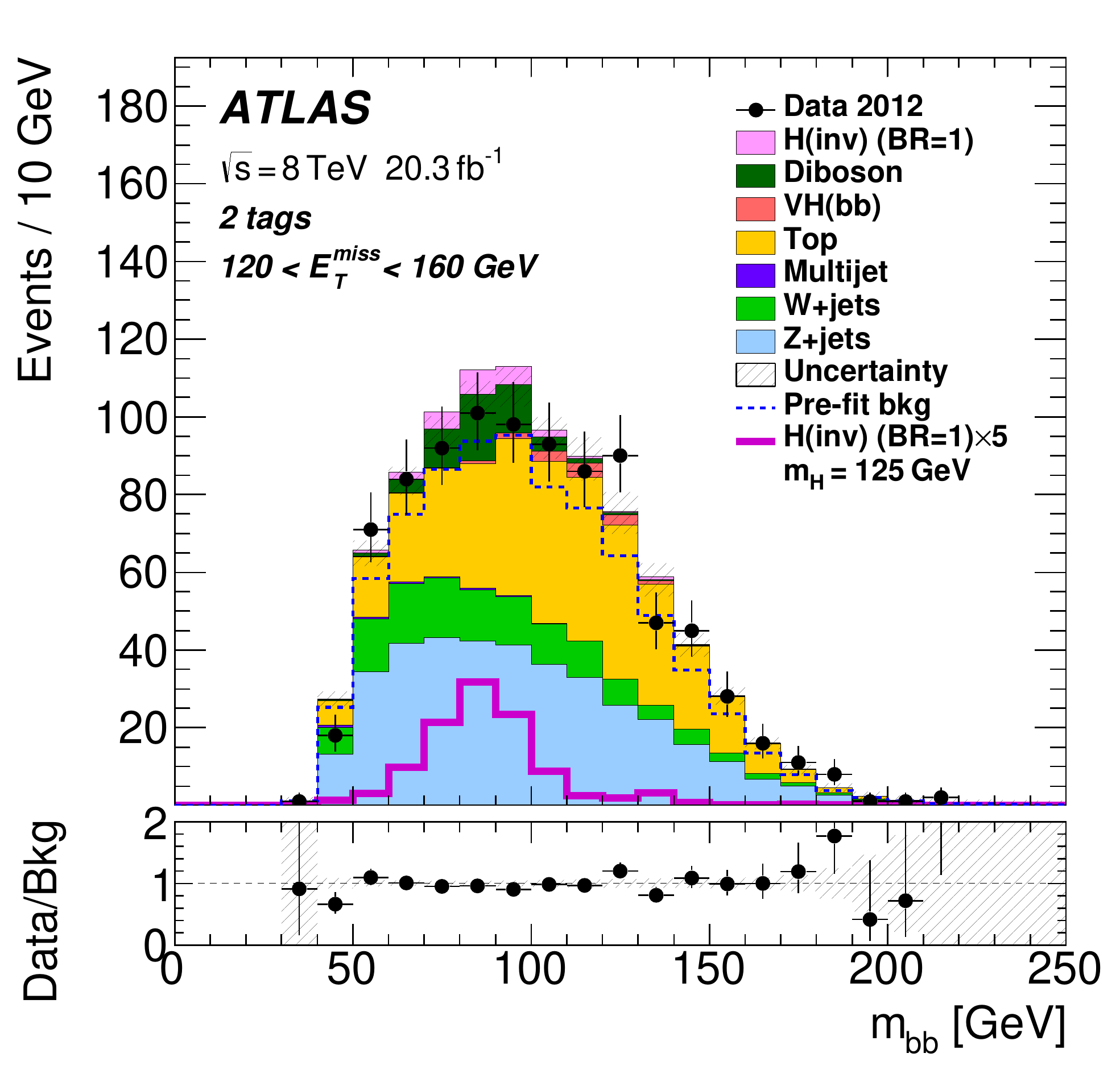}
}
\subfigure[]{
	\includegraphics[width=0.48\textwidth]{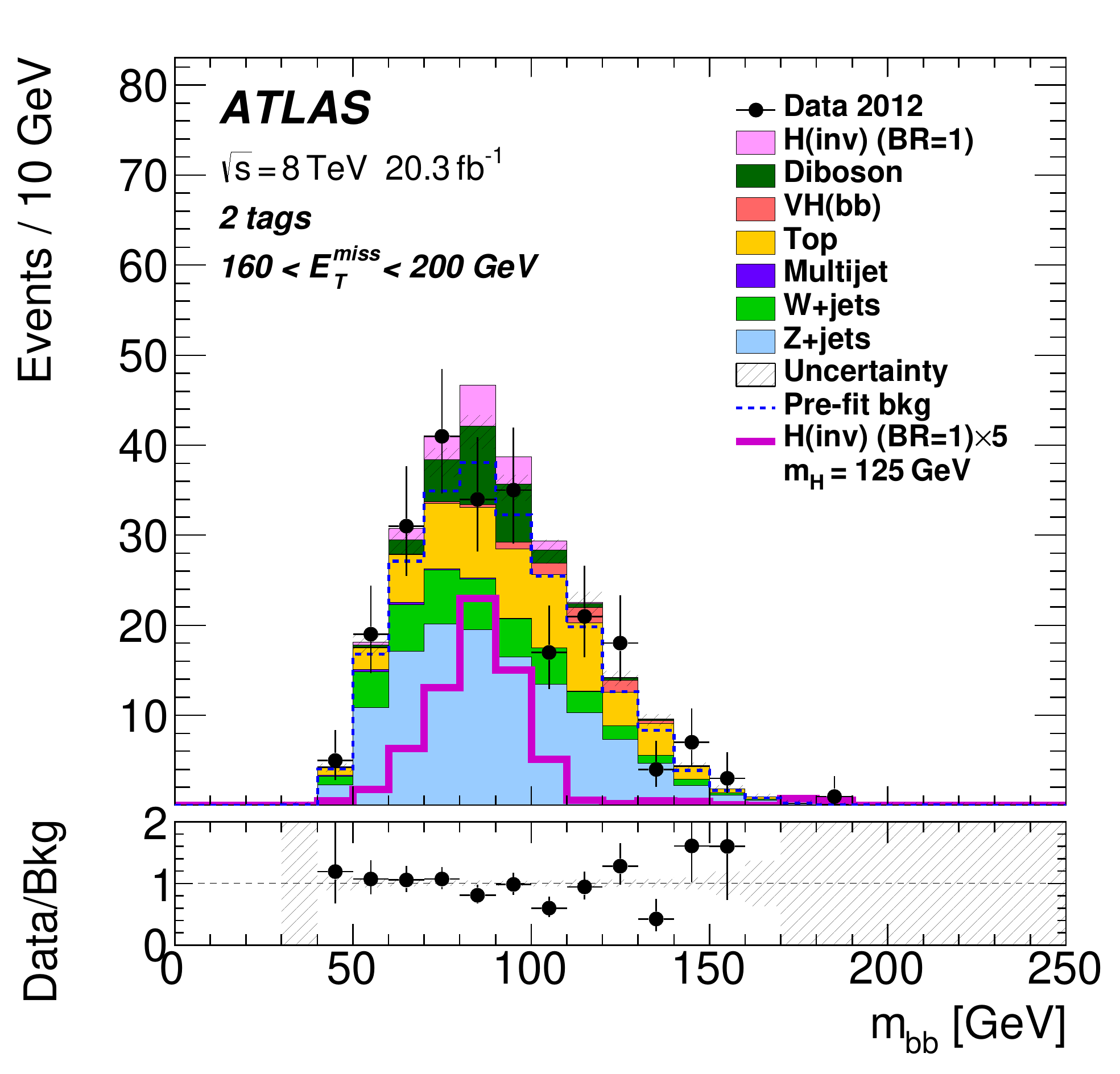}
}
\subfigure[]{
	\includegraphics[width=0.48\textwidth]{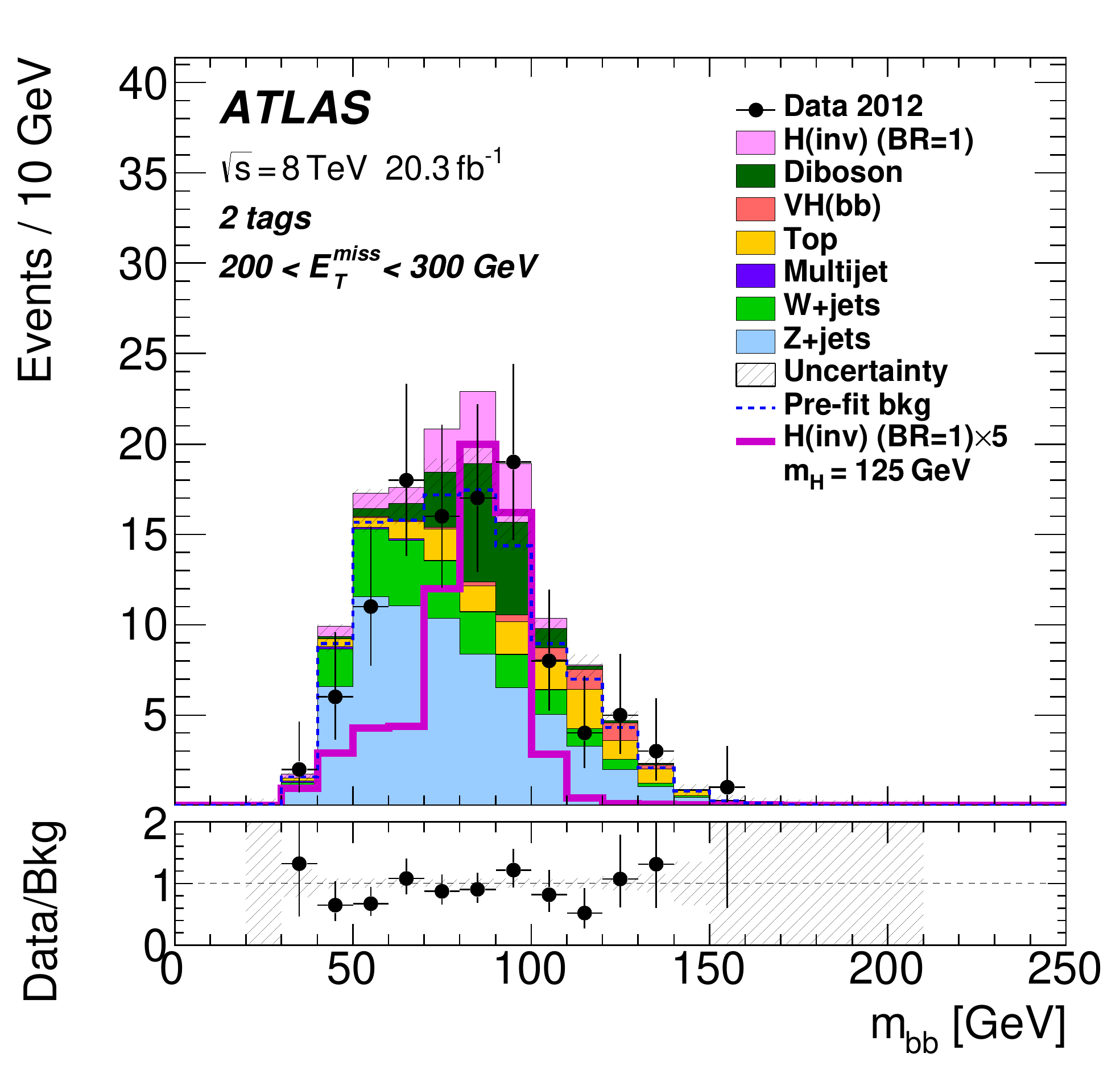}
}
\subfigure[]{
	\includegraphics[width=0.48\textwidth]{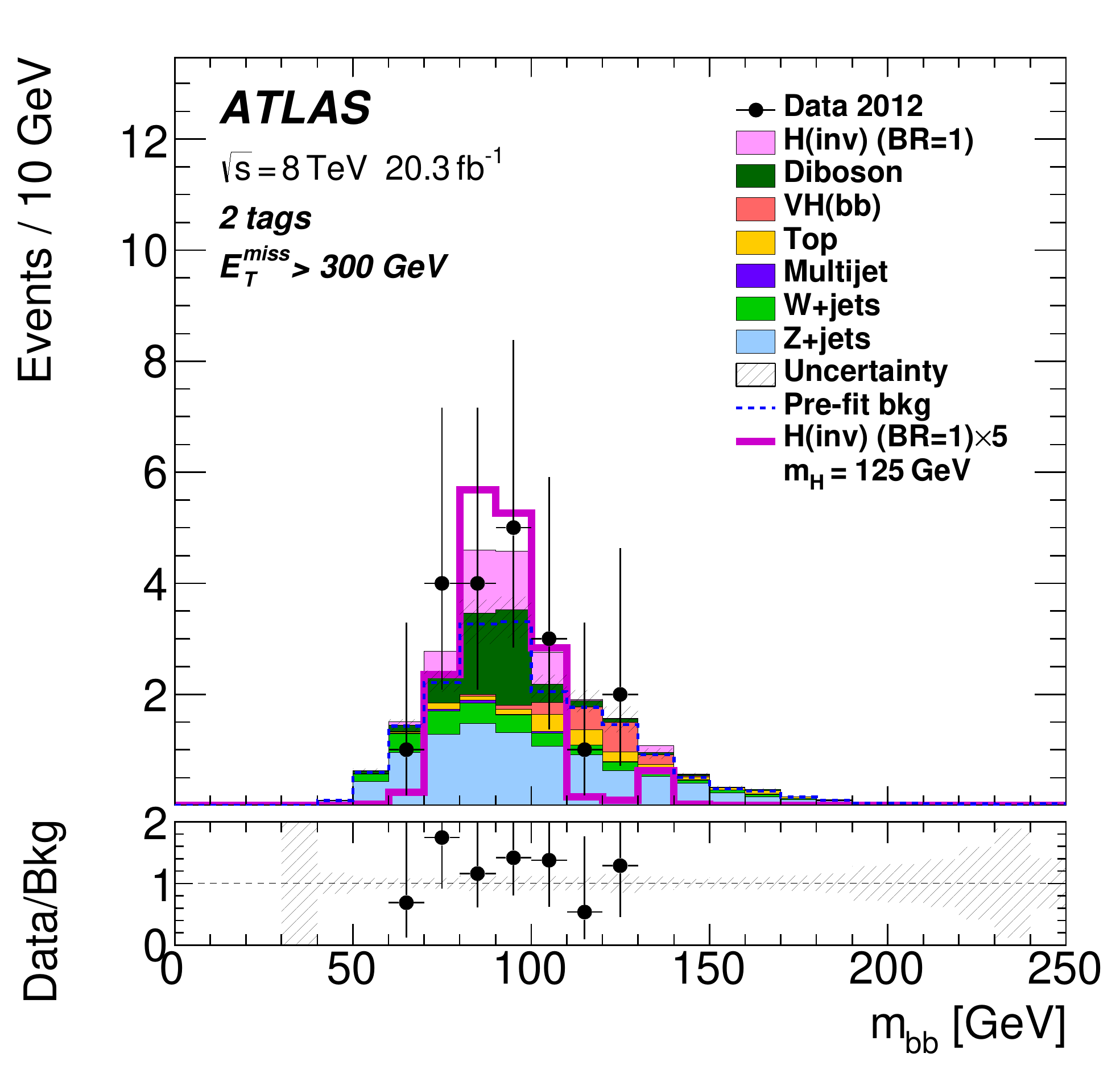}
}
\caption{
The dijet invariant mass ($m_{bb}$) distributions in
the signal region for the 2-$b$-tag category, for events with \met\ in the range
(a) [120--160 GeV], (b) [160--200 GeV], (c) [200--300 GeV] and (d) [$>$ 300 GeV].
The data are compared with the background model after the likelihood fit.
The bottom plots show the ratio of the data to the total background.
The signal expectation for $m_H = 125\,$GeV is shown on top of the
background and additionally as an overlay line, scaled by the factor
indicated in the legend.
The total background before the fit is shown as a dashed line.
The hatched bands represent the total uncertainty on the background.
}
\label{fig:postfit_mass_2tag}
\end{figure*}

The null results are used to set 95\% confidence level (CL) upper limits
on the product of the $VH$ cross sections and the $V \rightarrow jj$ and
$\Hinv$ decay branching ratio,
$\sigma_{VH} \times \text{BR}(\Hinv)$,
as a function of the Higgs boson mass
in the range $115<m_H<300$~GeV as shown in Fig.~\ref{fig:VH_XS_limits}.
The limits are computed with a modified frequentist method, also known as CL$_s$~\cite{Read:2002hq}, and a profile-likelihood-based test statistic~\cite{Cowan:2010js}.
At $m_H=125$~GeV, for $VH$ production, a limit of 1.1~pb is observed compared with 1.1~pb expected.
These combined results for $VH$ production assume the SM proportions of the $WH$ and $ZH$ contributions.
Observed (expected) limits are also derived for the two contributions separately,
1.2 (1.3)~pb for $WH$ and 0.72 (0.59)~pb for $ZH$.
As shown in Table~\ref{tab:event_yields}, the 2-tag categories are almost only sensitive to $ZH$, 
the 1-tag categories are equally sensitive to $WH$ and $ZH$, and the 0-tag categories are more sensitive to $WH$ production.
The two processes contribute approximately equally to the sensitivity.

\begin{figure*}
\centering
\includegraphics[width=0.6\textwidth]{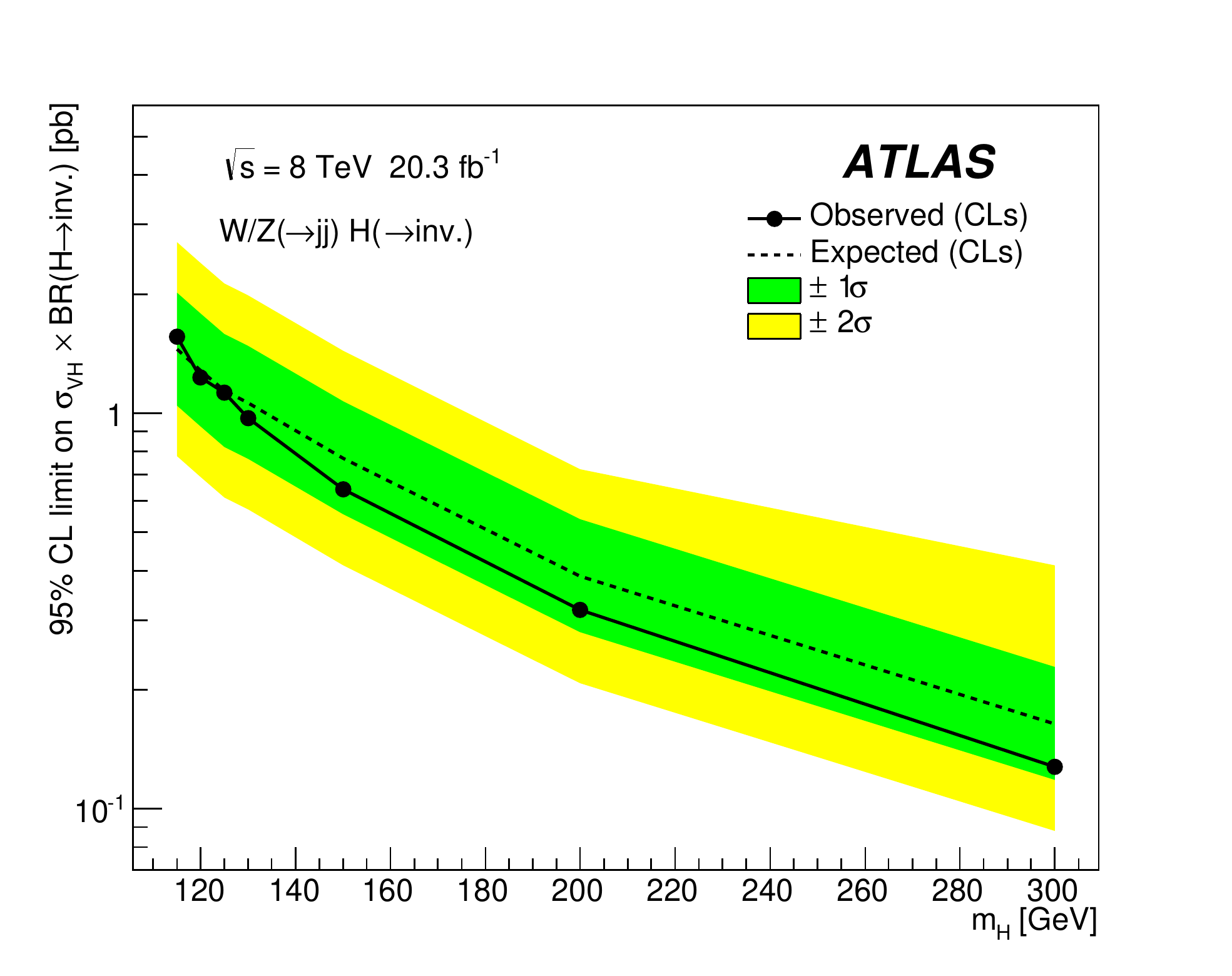}
\caption{\label{fig:VH_XS_limits}
Upper limits on $\sigma_{VH} \times \text{BR}(\Hinv)$
at 95\% CL for a Higgs boson with
$115 <m_H< 300$~GeV. The full and dashed lines show the observed and expected limits, respectively.
}
\end{figure*}

For the discovered Higgs boson at $m_H=125$~GeV,
an observed (expected) upper limit of 78\% (86\%) at 95\%~CL on the
branching ratio of the Higgs boson to invisible particles is set.
These limits are derived assuming SM production and combining contributions
from $VH$ and gluon-fusion processes.
The gluon-fusion production process contributes about 39\% (29\%) to
the observed (expected) combined sensitivity.

\FloatBarrier
\section{Summary}
\label{sec:summary}
In summary, Higgs boson decays to particles that are invisible to
the ATLAS detector are searched for in the final states of two or three jets
and large missing transverse momentum in a $pp$ collision dataset corresponding
to an integrated luminosity of 20.3~\ifb\ at a centre-of-mass energy of 8~TeV.
No excess of events over the expected backgrounds is observed.
The results are used to constrain the cross section for $VH$ production
followed by the decay $\Hinv$ for $115<m_H<300$~GeV.
The observed 95\% CL upper limit on
$\sigma_{VH} \times \text{BR}(\Hinv)$ varies from
1.6~pb at 115~GeV to 0.13~pb at 300~GeV.
Assuming SM production and including the $gg\to H$ contribution,
an observed (expected) upper limit of 78\% (86\%) on BR$(\Hinv)$ is derived
for the discovered Higgs boson with $m_H=125$~GeV.
This independent result is comparable to that of the ATLAS $ZH$ search
with $Z\to\ell\ell$ and $\Hinv$~\cite{Aad:2014iia}.

\section*{Acknowledgements}

We thank CERN for the very successful operation of the LHC, as well as the
support staff from our institutions without whom ATLAS could not be
operated efficiently.

We acknowledge the support of ANPCyT, Argentina; YerPhI, Armenia; ARC,
Australia; BMWFW and FWF, Austria; ANAS, Azerbaijan; SSTC, Belarus; CNPq and FAPESP,
Brazil; NSERC, NRC and CFI, Canada; CERN; CONICYT, Chile; CAS, MOST and NSFC,
China; COLCIENCIAS, Colombia; MSMT CR, MPO CR and VSC CR, Czech Republic;
DNRF, DNSRC and Lundbeck Foundation, Denmark; EPLANET, ERC and NSRF, European Union;
IN2P3-CNRS, CEA-DSM/IRFU, France; GNSF, Georgia; BMBF, DFG, HGF, MPG and AvH
Foundation, Germany; GSRT and NSRF, Greece; RGC, Hong Kong SAR, China; ISF, MINERVA, GIF, I-CORE and Benoziyo Center, Israel; INFN, Italy; MEXT and JSPS, Japan; CNRST, Morocco; FOM and NWO, Netherlands; BRF and RCN, Norway; MNiSW and NCN, Poland; GRICES and FCT, Portugal; MNE/IFA, Romania; MES of Russia and NRC KI, Russian Federation; JINR; MSTD,
Serbia; MSSR, Slovakia; ARRS and MIZ\v{S}, Slovenia; DST/NRF, South Africa;
MINECO, Spain; SRC and Wallenberg Foundation, Sweden; SER, SNSF and Cantons of
Bern and Geneva, Switzerland; NSC, Taiwan; TAEK, Turkey; STFC, the Royal
Society and Leverhulme Trust, United Kingdom; DOE and NSF, United States of
America.

The crucial computing support from all WLCG partners is acknowledged
gratefully, in particular from CERN and the ATLAS Tier-1 facilities at
TRIUMF (Canada), NDGF (Denmark, Norway, Sweden), CC-IN2P3 (France),
KIT/GridKA (Germany), INFN-CNAF (Italy), NL-T1 (Netherlands), PIC (Spain),
ASGC (Taiwan), RAL (UK) and BNL (USA) and in the Tier-2 facilities
worldwide.

\newpage

\printbibliography

\newpage 
\begin{flushleft}
{\Large The ATLAS Collaboration}

\bigskip

G.~Aad$^{\rm 85}$,
B.~Abbott$^{\rm 113}$,
J.~Abdallah$^{\rm 151}$,
O.~Abdinov$^{\rm 11}$,
R.~Aben$^{\rm 107}$,
M.~Abolins$^{\rm 90}$,
O.S.~AbouZeid$^{\rm 158}$,
H.~Abramowicz$^{\rm 153}$,
H.~Abreu$^{\rm 152}$,
R.~Abreu$^{\rm 30}$
% ...
\bigskip
\\
% ...
$^{11}$ Institute of Physics, Azerbaijan Academy of Sciences, Baku, Azerbaijan\\
% ...
$^{30}$ CERN, Geneva, Switzerland\\
% ...
$^{85}$ CPPM, Aix-Marseille Universit{\'e} and CNRS/IN2P3, Marseille, France\\
% ...
$^{90}$ Department of Physics and Astronomy, Michigan State University, East Lansing MI, United States of America\\
% ...
$^{107}$ Nikhef National Institute for Subatomic Physics and University of Amsterdam, Amsterdam, Netherlands\\
% ...
$^{113}$ Homer L. Dodge Department of Physics and Astronomy, University of Oklahoma, Norman OK, United States of America\\
% ...
$^{151}$ Institute of Physics, Academia Sinica, Taipei, Taiwan\\
$^{152}$ Department of Physics, Technion: Israel Institute of Technology, Haifa, Israel\\
$^{153}$ Raymond and Beverly Sackler School of Physics and Astronomy, Tel Aviv University, Tel Aviv, Israel\\
% ...
$^{158}$ Department of Physics, University of Toronto, Toronto ON, Canada\\
% ...
\end{flushleft}

\end{document}